\documentclass[10pt,journal]{IEEEtran}
\usepackage[top=0.48in, bottom=0.48in, left=0.48in, right=0.48in]{geometry} 
\IEEEoverridecommandlockouts
\usepackage{cite}
\usepackage{epsfig}
\usepackage{tikz}
\usepackage{lettrine}
\usepackage{enumitem}
\usepackage{dirtytalk}
\usepackage[table]{xcolor}
\usepackage{amsthm}
\newtheorem{remark}{Remark}
\newtheorem{lemma}{Lemma}
\usepackage{amsmath,amssymb,amsfonts}
\allowdisplaybreaks
\usepackage{mathrsfs}
\usepackage{algorithmic}
\usepackage{graphicx}
\usepackage{subcaption}
\usepackage{textcomp}
\usepackage{xcolor}
\usepackage{amsmath}
\usepackage{cuted}
\usepackage{xparse}
\usepackage{amsfonts}
\usepackage[multiple]{footmisc}
\usepackage{amssymb}
\usepackage{mathtools}
\usepackage{stackengine}
\usepackage[mathscr]{euscript}
\usepackage{graphics}
\usepackage[T1]{fontenc}
\usepackage{verbatim} 
\usepackage{multicol}
\usepackage{svg}
\usepackage{lettrine}
\usepackage[table]{xcolor}
\usepackage{etoolbox}
\allowdisplaybreaks
\usepackage{mathrsfs}
\usepackage{afterpage}
\usepackage[font=footnotesize,labelfont=footnotesize]{caption}
\usepackage[ruled,vlined]{algorithm2e}
\usepackage{listings}
\usepackage{url}
\usepackage{pdfpages}
\usepackage{float}
\usepackage{mathptmx}
\usepackage[11pt]{moresize}
\usepackage{stfloats}
\usepackage{diagbox}
\usepackage{wrapfig}
\usepackage{balance}
\allowdisplaybreaks
\usepackage[caption=false,font=footnotesize]{subfig}
\makeatother
\title{Pinching Antenna Assisted Integrated Sensing and Communication Using Quadrature RSMA}

\author{Sagnik Bhattacharyya, ~\IEEEmembership{Member, IEEE}, Shiv Kumar, ~\IEEEmembership{Member, IEEE},\\ Keshav Singh, ~\IEEEmembership{Senior Member, IEEE}, and Chih-Peng Li,~\IEEEmembership{Fellow,~IEEE}

\thanks{Sagnik Bhattacharyya, Keshav Singh, and Chih-Peng Li are with the Institute of Communications Engineering, National Sun Yat-sen University, Kaohsiung 80424, Taiwan (e-mail: sagnikbhattacharyya95@gmail.com, keshav.singh@mail.nsysu.edu.tw, cpli@faculty.nsysu.edu.tw).}

\thanks{Shiv Kumar is with the Department of Electrical Engineering, Indian Institute of Technology Delhi (e-mail: shiv.iitrpr@gmail.com).}

\vspace{-2.5em}
}
\begin{document}
\maketitle
\begin{abstract}
    This paper proposes a pinching antenna system (PASS)-assisted downlink quadrature rate-splitting multiple access (Q-RSMA)-based integrated sensing and communication (PASS-QRSMA-ISAC) framework for simultaneously supporting multi-user communication and target sensing. In the proposed framework, multiple pinching antennas (PAs) deployed along a dielectric waveguide serve communication users (CUs) distributed across a dual-room indoor environment, where line-of-sight (LoS) and non-line-of-sight (NLoS) propagation conditions coexist. By integrating PASS with Q-RSMA, the proposed framework improves interference management while reducing the dependence on successive interference cancellation. The communication and sensing signal models are developed, followed by the corresponding SNR/SINR expressions. To obtain tractable analytical insights, a two-user single-PA case under ideal waveguide conditions is first analysed, where the per-user communication outage probability (COP), system COP, sensing outage probability (SOP), and ergodic sum-rate are formulated using distance statistics, CDF/CCDF characterisations, and numerical integration methods. The analysis is then extended to the generalised multi-user multi-PA scenario by developing statistical characterisations for the LoS communication, NLoS communication, and sensing channels. Based on these characterisations, the per-user COP, system COP, SOP, and ergodic sum-rate are formulated and evaluated for the generalised framework. Monte Carlo simulations validate the analytical results and demonstrate the performance gains of the proposed PASS-QRSMA-ISAC framework over the considered benchmark schemes, including PASS-RSMA-ISAC, PASS-NOMA-ISAC, and PASS-SDMA-ISAC. The results further highlight the effects of various system parameters on the proposed framework.
\end{abstract}
\vspace{-1em}
\begin{IEEEkeywords}
Pinching Antenna, Rate Splitting Multiple Access (RSMA), Quadrature Rate Splitting Multiple Access (Q-RSMA), Integrated Sensing and Communication (ISAC). 
\end{IEEEkeywords}
\vspace{-1.25em}
\section{Introduction}
\lettrine{T}{he} growing demand for ultra-high data rates is driving wireless communication systems toward the millimeter-wave (mmWave) and terahertz (THz) bands, which offer significantly larger bandwidths. Notably, these high-frequency spectra increasingly overlap with those traditionally allocated for radar sensing, while emerging applications such as autonomous driving, collision avoidance, surveillance, robotics, virtual reality (VR), and massive Internet-of-Things (IoT) deployments require both reliable information exchange and high-resolution environmental perception \cite{10054381}. This motivates the integration of communication and sensing functionalities within a unified platform. However, although mmWave and THz frequencies provide abundant bandwidth resources, integrating sensing and communication within the same platform introduces several challenges, including co-channel interference, increased feedback and computational complexity, and additional hardware overhead when separate transmitters and processing chains are employed \cite{8828016}. To address these limitations, integrated sensing and communication (ISAC) has emerged as a promising paradigm that enables both functionalities to share the same spectral, hardware, and signal-processing resources, thereby improving spectrum efficiency, reducing system complexity, and mitigating interference \cite{9737357}. Moreover, ISAC has recently attracted significant attention as a key enabling technology for future sixth-generation (6G) wireless networks, supporting advanced applications such as smart cities, autonomous driving, and intelligent transportation \cite{ShivSir}. \textit{Despite these advantages, ISAC operation in the mmWave/THz bands remains challenging due to severe path loss, line-of-sight (LoS) blockage, and deep shadowing, which can significantly degrade reliability even over short or moderate transmission distances \cite{6732923}.}
\par To address these issues, recent studies have explored the concept of actively reconfiguring the wireless propagation environment using flexible antenna technologies, including reconfigurable intelligent surfaces (RISs) \cite{10781415}, fluid antennas (FAs) \cite{9264694}, and movable antennas (MAs) \cite{10318061}. RISs consist of programmable reflecting elements capable of imposing controllable phase shifts on incident electromagnetic waves, thereby reshaping the wireless channel to extend coverage, bypass obstacles, and mitigate interference \cite{9326394,ShivWCL}. Meanwhile, FAs employ conductive liquids enclosed within dielectric media to dynamically modify antenna positions and radiation characteristics, enabling beam steering and diversity gains \cite{9264694}. Similarly, MAs mechanically reposition antenna elements to adapt to varying channel conditions \cite{10318061}. Consequently, these technologies have also found applications in ISAC systems \cite{10781415,10870338,10707252}. \textit{Nevertheless, both FAs and MAs are typically constrained to movements within only a few wavelengths, which results in negligible influence on large-scale path loss \cite{ShivSir}. Furthermore, once deployed, the number of active antennas in such systems cannot be easily reconfigured.}
\par To overcome the aforementioned limitations, pinching-antenna systems (PASS) have recently emerged as a promising architecture for flexible antenna deployment \cite{suzuki2022pinching}. In PASS, electromagnetic signals propagate through a dielectric waveguide, where small discrete pinching elements can be selectively activated to radiate the guided signal into free space. This architecture enables flexible and on-demand antenna deployment without physically relocating the antenna elements. By controlling the positions of the pinching elements, PASS can establish favourable LoS links with the intended users, thereby improving propagation conditions and mitigating the impact of large-scale path loss. Owing to these advantages, several recent studies have investigated the performance and design aspects of PASS \cite{10945421,10896748,10976621,10981775}. The study in \cite{10896748} focused on the influence of antenna positioning on key channel characteristics such as channel gain, path loss, and phase variations, and proposed placement strategies aimed at enhancing downlink transmission efficiency and improving signal quality at the receiver. In \cite{10976621}, analytical expressions for the outage probability and ergodic rate were developed for PASS-based systems, revealing that attenuation within the waveguide can have a noticeable impact on system performance, particularly for longer waveguide structures. In another work, authors in \cite{10981775} examined the array gain properties of PASS and showed that careful optimization of both the number of antennas and their spatial spacing can lead to significant improvements in overall system performance. Furthermore, beyond conventional communication applications, PASS has also recently attracted attention for its potential use in ISAC systems \cite{ding2025pinching,11122551,11197530,11414134,11212813}. In this context, \cite{ding2025pinching} derived closed-form Cramér–Rao lower bounds (CRLBs) for PASS-assisted ISAC (PASS-ISAC) systems and demonstrated a significant reduction in the CRLB compared to conventional architectures. In \cite{11122551}, a maximum-entropy reinforcement learning strategy was proposed to maximize the sum communication rate while satisfying radar signal-to-noise ratio (SNR) requirements. Furthermore, authors in \cite{11197530} investigated a PASS-ISAC system employing separate transmit and receive waveguides and optimized the illumination power for sensing while maintaining communication constraints.
\par On the other hand, advanced multiple access (MA) schemes such as space division multiple access (SDMA), non-orthogonal multiple access (NOMA), and rate-splitting multiple access (RSMA) have been developed to efficiently share radio resources among multiple users and manage multi-user interference. Among these, RSMA has recently emerged as a generalized MA framework that bridges the gap between SDMA and NOMA by enabling partial interference decoding while treating the remaining interference as noise \cite{7470942,9663192,mao2018rate}. The concept of 1-layer RSMA for downlink multi-user MIMO was introduced in \cite{7470942}, while \cite{9663192} further investigated its information- and communication-theoretic performance with emphasis on precoder optimization and practical PHY-layer design. In RSMA, each user message is divided into a common stream decoded by all users and private streams intended for individual users, which provides flexible interference management and has been shown to achieve superior performance compared to SDMA and NOMA across various network conditions \cite{mao2018rate}. Owing to these advantages, RSMA has attracted significant research attention in recent years \cite{bansal2022rate,11193931,10906512,11367373,10880921}. Motivated by its strong interference management capability, the integration of RSMA with ISAC systems has also gained increasing interest \cite{10847905,11364301,11365995}. Furthermore, by leveraging the complementary advantages of RSMA and PASS, recent studies have investigated PASS-assisted RSMA frameworks to exploit the flexible antenna deployment offered by PASS for improving communication performance \cite{11304137}. In addition, motivated by the growing interest in PASS-assisted ISAC systems \cite{ding2025pinching,11122551,11197530,11414134,11212813}, the authors in \cite{ShivSir} recently proposed a PASS-RSMA-ISAC framework to enhance both communication reliability and sensing capability. In that work, PASS was employed to flexibly control the wireless propagation environment, while the ISAC framework enabled the simultaneous execution of communication and sensing functionalities. Moreover, RSMA was utilized to efficiently manage inter-user interference during communication. \textit{However, \cite{ShivSir} primarily focused on outage analysis for a two-user single-PA scenario, while the outage and sum-rate performance of the multi-user multi-PA scenario remains unexplored. Extending the analysis to the multi-PA case is non-trivial, since the effective PASS channel depends on the superposition of multiple PA-level links, including in-waveguide attenuation, phase variation, PA-user geometry, and heterogeneous LoS/NLoS wireless propagation. Furthermore, most existing RSMA-based systems adopt the conventional one-layer RSMA structure, which relies on successive interference cancellation (SIC) to separate the common and private streams. As a result, their performance becomes susceptible to error propagation caused by imperfect SIC (ipSIC), which can significantly degrade system reliability in practical scenarios where perfect interference cancellation cannot be guaranteed.}
\par Motivated by this limitation, the authors in \cite{9831449} recently proposed a zero-SIC-based rate-splitting framework referred to as quadrature rate-splitting multiple access (Q-RSMA). Similar to conventional 1-layer RSMA, the user messages in Q-RSMA are first split to form common and private streams. However, unlike conventional RSMA where both streams are transmitted over the same signal dimension and separated using SIC, Q-RSMA maps the common stream onto the in-phase component of the signal, while the combined private streams are mapped onto the quadrature component \cite{9831449}. Consequently, the common stream becomes inherently free from interference caused by the private streams, enabling more reliable decoding. Thus, this signal-domain separation enhances the reliability of common-stream decoding and reduces the reliance on SIC. This property is particularly relevant in ISAC systems, where the communication streams are already affected by sensing-signal interference. Therefore, avoiding SIC-based stream separation helps prevent an additional residual-interference component from degrading the communication decoding process. However, this benefit is achieved at the cost of reduced degrees of freedom (DoF) for each stream, since Q-RSMA maps the common and private streams onto separate real-valued signal components. This trade-off is reflected later through a rate-scaling factor in the achievable-rate expressions. Despite this trade-off, Q-RSMA remains attractive due to its reduced reliance on SIC and improved robustness against SIC-induced error propagation, which have motivated several recent investigations on Q-RSMA-based communication systems \cite{10898709,10920858,TVT_QRSMUC}. \textit{Nevertheless, although these properties make Q-RSMA promising for ISAC scenarios, its application to ISAC frameworks remains largely unexplored. Furthermore, to the best of the authors' knowledge, the joint integration of Q-RSMA with PASS in ISAC-enabled wireless networks has not been investigated in the existing literature.}
\par Summarising the above discussion, existing works have mainly analysed two-user single-PA based PASS-ISAC systems, while, to the best of the authors’ knowledge, the performance analysis of multi-user multi-PA networks remains unexplored. Moreover, Q-RSMA is particularly relevant for future QoS-demanding ISAC networks, as its signal-domain separation reduces the dependence on SIC and improves communication reliability under sensing-signal interference. Motivated by this, we propose a PASS-assisted downlink Q-RSMA-based ISAC (PASS-QRSMA-ISAC) network, where multiple PAs deployed along a waveguide simultaneously support multi-user communication and target sensing. The \textbf{main contributions} of this paper can be listed as follows:
\begin{itemize}
    \item \textbf{Proposed PASS-QRSMA-ISAC framework:} We propose a generalised PASS-assisted downlink Q-RSMA-based ISAC framework, where multiple PAs deployed along a dielectric waveguide are used to simultaneously serve multiple communication users (CUs) and perform target sensing. The CUs are flexibly distributed across two adjacent rooms, thereby capturing heterogeneous indoor propagation conditions with both line-of-sight (LoS) and non-line-of-sight (NLoS) links. Based on the proposed architecture, the generalised received signal models for communication and sensing are developed, followed by the corresponding signal-to-noise ratio (SNR)/ signal-to-interference-plus-noise ratio (SINR) expressions.
    \item \textbf{Two-User Single-PA Performance Analysis:} To obtain tractable analytical insights, we first consider a two-user single-PA special case under an ideal waveguide scenario and derive the corresponding outage and sum-rate expressions.
    \begin{itemize}
        \item[(a)] \textbf{Outage Analysis:} The communication outage probability (COP) expressions are derived using the distance statistics of the LoS and NLoS users. For the LoS user, a piecewise closed-form CDF of the squared PA-to-user distance is used, whereas for the NLoS user, a piecewise PDF-based formulation is evaluated using Gaussian-Chebyshev quadrature. The system COP is obtained from the individual user outage events. The mean-squared error (MSE)-based sensing outage probability (SOP) is derived by mapping the CRLB-based sensing accuracy requirement to an equivalent sensing-SINR threshold and applying the piecewise closed-form CDF of the squared PA-to-sensing-target distance.
        \item[(b)] \textbf{Ergodic Sum Rate Analysis:} The ergodic sum-rate is formulated using the derived CDF/CCDF characterisations for the LoS and NLoS CUs, while accounting for the common-rate and private-rate components of Q-RSMA. The final expression is evaluated using numerical integration methods.
    \end{itemize}
    \begin{table}[t]
\centering
\caption{List of Notations and Mathematical Operators.}
\label{Notation}
\begin{tabular}{|>{\centering\arraybackslash}m{2.25cm}||p{5.75cm}|}
 \hline
 \rowcolor{green!20}\textbf{Symbols} &\hspace{20mm} \textbf{Description}\\
 \hline
  \rowcolor{blue!20}\multicolumn{2}{|c|}{\textbf{General System Parameters}} \\
  \hline
  $K$, $M$, $K-M$ & Total number of CUs, number of CUs in Room 1 and number of CUs in Room 2, respectively\\
  $L$ & Total number of PAs\\
 $\EuScript{P}_t$ & Total transmit power\\
 $\alpha$, $(1-\alpha)$ & Power allocation coefficient between the communication signal and sensing signal, respectively\\
 $h_{\omega,l}, \textbf{h}_\omega$ & In-waveguide channel between $l^{th}$ PA and feed point, and In-waveguide channel vector, respectively\\
 $D$, $d$ & Length/ width of a room and the height at which waveguide is deployed, respectively\\
 $\eta_{\text{eff}}$ & Effective refractive index\\
 $\kappa$ & In-waveguide attenuation coefficient\\
 $c$, $f_C$ & Speed of light and carrier frequency, respectively\\
 $\Phi_F=[x_F, 0, d]$ & Coordinate of the feed point\\
 $\Phi_{P_l}=\left[x_{P_l}, 0, d\right]$ & Coordinate of $l^{th}$ PA\\
 \hline
 \rowcolor{blue!20}\multicolumn{2}{|c|}{\textbf{Communication Task Parameters}} \\
 \hline
 $U_k$ & $k^{th}$ CU\\
 $h_{k,l}, \textbf{h}_k$ & Wireless channel between $U_k$ and $l^{th}$ PA, and wireless channel vector between $U_k$ and all PAs, respectively\\
 $z_k$ & Background AWGN at $U_k$\\
 $\nu$ & Path loss exponent for user in Room 2\\
 $\beta_\text{C}$, $\beta_{\text{P}i}$ & Power allocation coefficients of the common and $i^{th}$ private stream, respectively\\
 $\Lambda_{k,Com,Pin}$, $\Lambda_{k,Pvt,Pin}$ & SNR/ SINR of the common stream and private stream at $U_k$, respectively\\
 $\tau_{Ck}$, $\tau_{Pk}$ & Thresholds of the common stream and private stream at $U_k$, respectively\\
 $\EuScript{Y}_k$ & Signal received at $U_k$\\ 
 $\Phi_k=[x_k, y_k, 0]$ & Coordinate of $U_k$\\
 \hline
 \rowcolor{blue!20}\multicolumn{2}{|c|}{\textbf{Sensing Task Parameters}} \\
 \hline
 $U_S$ & Sensing target\\
 $h_{S,l}, \textbf{h}_S$ & Wireless sensing channel between $U_S$ and $l^{th}$ PA, and wireless sensing channel vector between $U_S$ and all PAs, respectively\\
 $z_S$ & Background AWGN at BS\\
 $\sigma_{RCS}$ & Radar Cross-Sectional Area\\
 $\Lambda_{Sensing,Pin}$ & SINR of the sensing signal\\
 $\tau_S$ & Sensing threshold\\
 $\EuScript{Y}_S$ & Sensing echo signal received at BS\\ 
 $\Delta_S$ & Residual interference caused by imperfect cancellation of the communication component during sensing-signal recovery at the BS\\
 $\Phi_S=[x_S, y_S, 0]$ & Coordinate of $U_S$\\
 \hline
 \rowcolor{blue!20}\multicolumn{2}{|c|}{\textbf{Mathematical Operators/ Symbols}} \\
 \hline
 $j$, $(.)^*$, $\lVert . \rVert$, $\lvert . \rvert$, $\mathbb{E}\{.\}$ & Imaginary unit $\left(\sqrt{-1}\right)$, Complex conjugate, Euclidean norm, absolute value and mean operators, respectively\\
 $\Re\{.\}$, $\Im\{.\}$ & Real and imaginary part of a complex signal/ stream, respectively\\
 $f_{RV}(.), \EuScript{F}_{RV}(.), F_{RV}(.)$& PDF, CDF, and CCDF of random variable RV\\
 \hline
\end{tabular}
\end{table}
    \item \textbf{Multi-User Multi-PA Performance Analysis:} The proposed model is generalised to a multi-user multi-PA setting, where multiple CUs are distributed across the two rooms and jointly served by multiple PAs. Since the resulting effective channel involves the superposition of multiple PA-level contributions with in-waveguide attenuation, phase variation, and user-dependent wireless propagation, we develop tractable statistical characterisations for the communication and sensing channels, which are then used for outage and sum-rate analyses.
    \begin{itemize}
        \item[(a)] \textbf{Statistical Channel Characterisation:} The effective communication and sensing channels are statistically characterised as follows:
        \begin{enumerate}
            \item \textbf{LoS Communication Channels:} For CUs in Room-1, the effective channel is approximated as a moment-matched complex Gaussian random variable with non-zero mean. Hence, the channel envelope follows a Rician distribution, and the channel power gain follows a non-central chi-square distribution.
            \item \textbf{NLoS Communication Channels:} For CUs in Room 2, a PA-conditioned semi-analytical characterisation is developed. Conditioned on the CU and PA locations, the effective channel power gain follows an exponential distribution. The Rayleigh fading is averaged in closed form, while the remaining spatial expectation is evaluated numerically.
            \item \textbf{Sensing Channels:} For the sensing link, the transmit-side channel is approximated using moment matching, while the echo-return path is characterised by conditioning on the ST location and selected receiving PA. The resulting SOP accounts for multi-PA sensing transmission and nearest-PA echo reception, and is evaluated using numerical integration.
        \end{enumerate}
        \item[(b)] \textbf{Outage and Sum-Rate Analysis:} Based on the above statistical characterisations, the per-user COP, system COP, SOP, and ergodic sum-rate are formulated for the generalised multi-user multi-PA framework. The final expressions are evaluated using numerical integration methods.
    \end{itemize}
    \item \textbf{Performance Validation and Insights:} Extensive numerical results are presented to validate the analytical expressions and evaluate the proposed PASS-QRSMA-ISAC framework across per-user COP, system COP, SOP, and ergodic sum-rate. The validation and insights are presented for the following cases:
    \begin{itemize}
        \item[(a)] \textbf{Two-User Single-PA Scenario:} The analytical expressions are validated through Monte Carlo simulations under the ideal waveguide scenario. The proposed framework is compared with conventional PASS-RSMA-ISAC, PASS-NOMA-ISAC, and PASS-SDMA-ISAC schemes.
        \item[(b)] \textbf{Generalised Multi-User Multi-PA Scenario:} The analytical expressions are validated for the multi-user multi-PA setting, and the proposed framework is compared with PASS-RSMA-ISAC. The results investigate the impact of user distribution, power allocation, waveguide attenuation, and sensing-related parameters on the communication and sensing performance.
    \end{itemize}
    The results confirm the accuracy of the analytical derivations and highlight the performance gains over the considered benchmark schemes.
\end{itemize}
\par \textit{\textbf{Notations Used:}} The notation and mathematical operators used in the work are listed in Table \ref{Notation}.
\par \textit{\textbf{Organization of the paper:}} The remainder of the paper is organised as follows. Section \ref{systemmodel} presents the system model and SINR formulations. Sections \ref{singlePA_analysis} and \ref{multiplePA_analysis} provide the outage and sum-rate analyses for the two-user, single-PA and multi-user, multi-PA scenarios, respectively. Section \ref{results} presents the analytical and simulation results, and Section \ref{conclusion} concludes the paper.
\section{System Model}
\label{systemmodel}
\begin{figure}[t]
    \centering
    \includegraphics[scale=0.25]{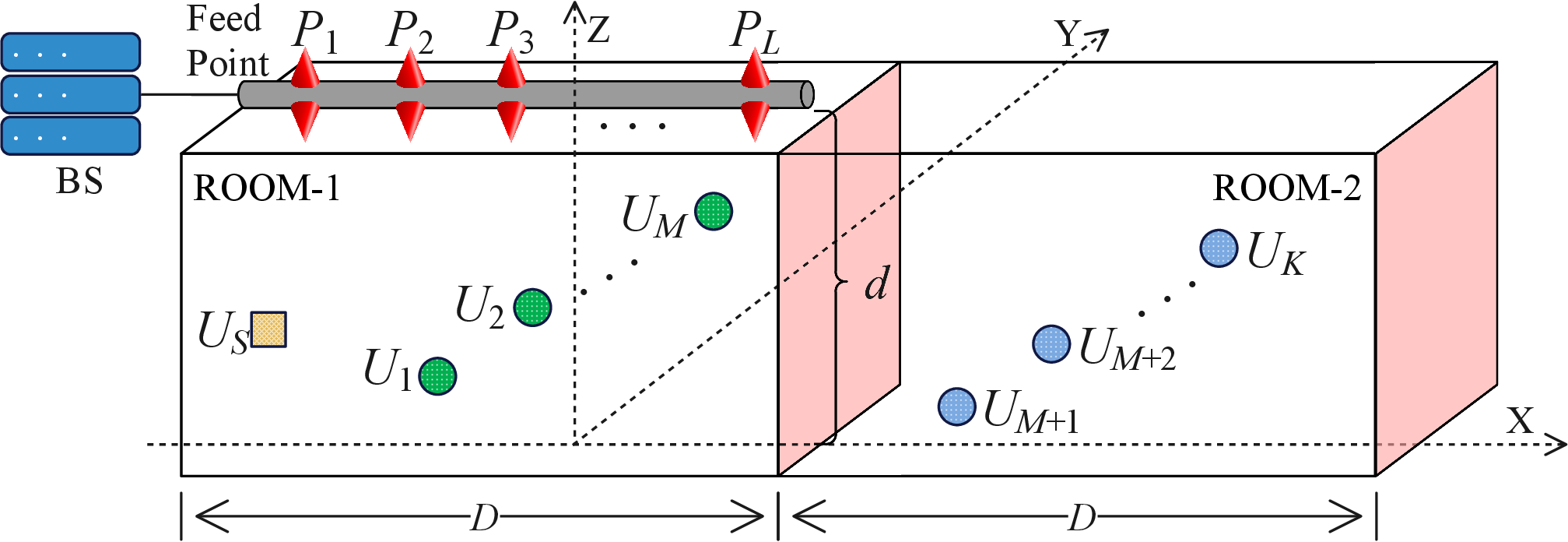}
    \caption{An illustration of multi-user multi-PA PASS-QRSMA-ISAC system.}
    \label{GeneralDiagram}
\end{figure}

Let us consider a PASS-assisted Q-RSMA-based downlink ISAC network comprising two adjacent rooms, each with dimensions $D \times D$, as illustrated in \figurename{~\ref{GeneralDiagram}}. A total of $K$ CUs are distributed across the two rooms, where $M$ CUs are located in Room 1 and the remaining $K-M$ CUs are located in Room 2. Any CU is denoted by $U_k$, where $k\in\{1,\ldots,K\}$. Without loss of generality, users $U_1,\ldots,U_M$ are located in Room 1, whereas the remaining $K-M$ CUs are located in Room 2. The coordinate of any CU $U_k$ is represented as $\Phi_k=[x_k,y_k,0]$. Accordingly, for users in Room 1, $x_k\in\left[-\frac{D}{2},\frac{D}{2}\right]$ and $y_k\in\left[-\frac{D}{2},\frac{D}{2}\right]$, where $k\leq M$. Similarly, for users in Room 2, $x_k\in\left[\frac{D}{2},\frac{3D}{2}\right]$ and $y_k\in\left[-\frac{D}{2},\frac{D}{2}\right]$, where $M+1\leq k\leq K$. Furthermore, a sensing target (ST), denoted by $U_S$, is located in Room 1. The coordinate of the ST is represented as $\Phi_S=[x_S,y_S,0]$, where $x_S\in\left[-\frac{D}{2},\frac{D}{2}\right]$ and $y_S\in\left[-\frac{D}{2},\frac{D}{2}\right]$. The users in the network are served by a waveguide deployed along the X-axis at a height $d$. The waveguide is equipped with $L$ PAs, denoted by $P_l \forall l \in \{1,\ldots,L\}$, where the coordinate of the $P_l$ is given by $\Phi_{P_l} = [x_{P_l},0,d]$ where $x_{P_l}\in\left[-\frac{D}{2},\frac{D}{2}\right]$. The waveguide is connected to the BS through a dedicated RF chain, which injects the RF signal into the waveguide via a feed point located at its front end. The coordinate of this feed point is denoted by $\Phi_F = [x_F,0,d]$. Throughout this work, the distance between any two points is calculated using the Euclidean norm. This setup establishes the basic network scenario, including the user and PA locations. Based on this, the next sub-section describes the channel modelling for the network.

\vspace{-0.5em}
\subsection{Channel Model}
The signal in the network propagates through both the waveguide and wireless medium, leading to the following channel categorisation:
\begin{itemize}
    \item \textbf{Inside Waveguide:} The in-waveguide component represents the portion of the signal that propagates along the waveguide from the feed point to the PAs. During this propagation, the signal experiences both attenuation and phase variation. The attenuation is caused by propagation loss inside the dielectric waveguide and is characterized by the attenuation coefficient $\kappa$, measured in dB/m. On the other hand, the phase variation occurs because the signal propagates inside the waveguide with a guided wavelength, which differs from the free-space wavelength due to the material and structural properties of the waveguide. This effect is characterized by the effective refractive index, $\eta_{\text{eff}}$, which determines the guided wavelength as $\lambda_G=\frac{\lambda}{\eta_{\text{eff}}}$, where $\lambda=\frac{c}{f_C}$ is the free-space wavelength, $c$ is the speed of light, and $f_C$ is the carrier frequency. Thus, the corresponding in-waveguide channel between the feed point and $l^{th}$ PA can be given as:
    \begin{equation}
    \begin{split}
    h_{\omega,l}=10^{-\frac{\kappa\left\lVert \Phi_{P_l}-\Phi_{F} \right\rVert}{20}}e^{-j\frac{2\pi}{\lambda}\left(\eta_{\text{eff}}\left\lVert \Phi_{P_l}-\Phi_F \right\rVert\right)},
    \end{split}
    \end{equation}
    where $\lVert \Phi_{P_l}-\Phi_F \rVert$ represents the Euclidean distance between $P_l$ and the feed point. Now, since multiple PAs are deployed along the same waveguide, the signal experiences different propagation phases and attenuation before reaching each PA, depending on its distance from the feed point. Thus, the in-waveguide channel vector can be given as:
    \begin{equation}
    \label{Waveguide_Channel_Vector}
        \textbf{h}_\omega=[h_{\omega,1}, h_{\omega,2}, \ldots, h_{\omega,L}]\in\mathbb{C}^{1\times L}.
    \end{equation}
    The in-waveguide channel vector in \eqref{Waveguide_Channel_Vector} is common to both the communication and sensing functionalities, since the transmitted signal first propagates from the feed point to the PAs before being radiated into the wireless medium. However, the sensing functionality involves an additional echo-return path. In this work, we consider a single receiving PA, denoted by $\widehat{P_l}$ for collecting the echo reflected from the ST. The collected echo is then guided back to the BS through the waveguide. Accordingly, the corresponding receive-side in-waveguide channel is denoted by
    \begin{equation}
        \widehat{h_{\omega,l}}=10^{-\frac{\kappa\left\lVert \Phi_{\widehat{P_l}}-\Phi_{F} \right\rVert}{20}}e^{-j\frac{2\pi}{\lambda}\left(\eta_{\text{eff}}\left\lVert \Phi_{\widehat{P_l}}-\Phi_F \right\rVert\right)}.
    \end{equation}
    \item \textbf{Wireless Channel:} Once radiated from the waveguide through the PAs, the signal propagates through the wireless medium towards the CUs and ST. Thus, the wireless propagation channel can be classified into the following cases:
    \begin{itemize}
        \item \textbf{Communication Links (Towards CUs):} This represents the channels between the PAs and the CUs. Since the waveguide is deployed in Room 1, the CUs in Room 1 experience strong LoS connections. In contrast, the CUs in Room 2 do not have direct visibility with the PAs and therefore experience a non-line-of-sight (NLoS) propagation condition. Accordingly, the channel between $l^{th}$ PA and any user $U_k$ can be given as:
        \begin{equation}
        \label{Individual_Channel}
        \begin{split}
        &h_{k,l}=
        \begin{cases}
        \frac{\xi e^{-j\frac{2\pi}{\lambda}\left(\left\lVert \Phi_{P_l}-\Phi_k \right\rVert\right)}}{\left\lVert \Phi_{P_l}-\Phi_k \right\rVert}, & k\in\{1,\ldots,M\}\\ \vspace{-2mm} \\
        \frac{g_{k,l}}{\sqrt{\left\lVert \Phi_{P_l}-\Phi_k \right\rVert^{\nu}}}, & k\in\{M+1,\ldots,K\}
        \end{cases}
        \end{split}
        \end{equation}
        where $g_{k,l}\in \mathbb{CN}\{0,1\}$ is the small scale fading between the $l^{th}$ PA and $U_k$, for $M+1\leq k\leq K$, and $\xi=\frac{\lambda}{4\pi}$ is the free space path loss at a reference distance of 1m. Here, $\lVert \Phi_{P_l}-\Phi_k\rVert$ represents the Euclidean distance between $l^{th}$ PA and $U_k$ and $\nu$ is the path loss exponent. Now, since there are multiple PAs, the channel vector between all the PAs and $U_k$ can be represented as: \vspace{-0.0em}
        \begin{equation}
        \label{CU_Channel_Vector}
            \textbf{h}_{k}=[h_{k,1}, h_{k,2}, \ldots,h_{k,L}]^T\in\mathbb{C}^{L\times 1}.
        \end{equation}
        Thus, using \eqref{Waveguide_Channel_Vector} and \eqref{CU_Channel_Vector}, the effective channel between PAs and any CU $U_k$ can be given as:
        \begin{equation}
        \label{CU_Channel}
        \begin{split}
        h_{\Sigma,k}&=\textbf{h}_\omega\textbf{h}_k=\sum\nolimits_{l=1}^{L}h_{\omega,l}h_{k,l}.
        \end{split}
        \end{equation}
        \item \textbf{Sensing Links (PA-to-ST and ST-to-PA Paths):} Unlike the communication links, the sensing link involves both the propagation of the transmitted signal from the PAs to the ST and the return propagation of the reflected echo from the ST to the receiving PA. Therefore, the sensing channel is modelled through two distinct propagation paths which are further discussed as follows:
        \begin{itemize}
            \item \textbf{PA-to-ST Path:} This path represents the propagation of the sensing signal from the transmit PAs to the ST. Since the ST is located in Room 1 and has direct visibility with the waveguide-mounted PAs, the PA-to-ST path is modelled as a dominant LoS link. Accordingly, the channel between $P_l$ and ST can be expressed as: \vspace{-0.5em}
            \begin{equation}
                h_{S,l}=\frac{e^{-j\frac{2\pi}{\lambda}\left(\left\lVert \Phi_{P_l}-\Phi_S \right\rVert\right)}}{\left\lVert \Phi_{P_l}-\Phi_S \right\rVert},
            \end{equation}
            where $\left\lVert \Phi_{P_l}-\Phi_S \right\rVert$ represents the distance between $P_l$ and the ST. Since multiple PAs are deployed, the PA-to-ST channel vector can be given as:
            \begin{equation}
            \label{Sensing_Channel_Vector}
                \textbf{h}_S=[h_{S,1}, h_{S,2},\ldots,h_{S,L}]^T\in\mathbb{C}^{L\times 1}.
            \end{equation}
            Thus, using \eqref{Waveguide_Channel_Vector} and \eqref{Sensing_Channel_Vector}, the effective transmit sensing channel between the PAs and ST can be expressed as:
            \begin{equation}
                \label{Tx_Sense_Channel}h_{\Sigma,S}=\textbf{h}_\omega\textbf{h}_S=\sum\nolimits_{l=1}^{L}h_{\omega,l}h_{S,l}.
            \end{equation}
            \item \textbf{ST-to-PA Path:} This path represents the echo-return propagation from the ST to the receiving PA, $\widehat{P_l}$ after the sensing signal is reflected by the target. The PA closest to the ST is selected and activated for echo reception. Similar to monostatic PASS-assisted ISAC sensing models, the sensing receiver is assumed to be co-located with the selected receiving PA, $\widehat{P_l}$ which collects the echo reflected from the ST before the received signal is guided back to the BS through the waveguide \cite{ShivSir}. Therefore, the echo-return link is modelled through the ST-to$\widehat{P_l}$ wireless path followed by the receive-side in-waveguide propagation. Accordingly, the selected receiving PA, $\widehat{P_l}$ can be expressed as $\widehat{P_l}=P_{\arg \min_{l\in\{1,\ldots,L\}}\left\lVert \Phi_{P_l}-\Phi_{S} \right\rVert}$. Thus, the corresponding coordinate of the selected PA can be given as $\Phi_{\widehat{P_l}}=\Phi_{P_{\arg \min_{l\in\{1,\ldots,L\}}\left\lVert \Phi_{P_l}-\Phi_{S} \right\rVert}}$. Now, since the ST is located in Room 1 and has direct visibility with the waveguide-mounted PAs, the ST-to-PA path is also modelled as a dominant LoS link. Accordingly, the wireless channel between the ST and $\widehat{P_l}$ can be expressed as:
            \begin{equation}
                \widehat{h_{S,l}}=\frac{e^{-j\frac{2\pi}{\lambda}\left(\left\lVert \Phi_{S}-\Phi_{\widehat{P_l}} \right\rVert\right)}}{\left\lVert \Phi_{S}-\Phi_{\widehat{P_l}} \right\rVert}.
            \end{equation}
            Thus, the effective channel while receiving echo signal can be given as:
            \begin{equation}
                \label{Rx_Echo_Channel}\widehat{h_{\Sigma,S}}=\widehat{h_{\omega,l}} \widehat{h_{S,l}}.
            \end{equation}
            Using \eqref{Tx_Sense_Channel} and \eqref{Rx_Echo_Channel}, the effective sensing channel can be given as: \vspace{-0.5em}
            \begin{equation}
            \label{overall_sensing_channel}
                h_{sen}=\sqrt{\xi_S}h_{\Sigma,S}\widehat{h_{\Sigma,S}},
            \end{equation} 
            where $\xi_S=\frac{\lambda^2\sigma_{RCS}}{(4\pi)^3}$. Here, $\sigma_{RCS}$ is the radar cross-sectional area.
        \end{itemize}
    \end{itemize}
\end{itemize}
In the following sub-section, we present the signal modelling of the proposed multi-PA assisted ISAC framework based on Q-RSMA. Before proceeding, the key assumptions considered in this work are summarized as follows: (i) the BS is assumed to have perfect knowledge of the channel state information (CSI) for both communication and sensing links, and (ii) the PA configuration is assumed to remain fixed within each transmission block. Consequently, the wireless channel can be treated as quasi-static over the duration of data transmission and sensing. This assumption is also consistent with existing PASS-assisted ISAC studies such as \cite{ShivSir}. The detailed method for channel estimation of PASS can be found in \cite{11018390}.

\vspace{-0.5em}
\subsection{Signal Model}
In this sub-section, we present the signal model of the proposed PASS–QRSMA–ISAC framework. For the communication functionality of the ISAC system, Q-RSMA is adopted. For each user $U_k$, the corresponding data, $\zeta_k$, is split into two components, namely a common part, $\zeta_{\text{C}k}$ and a private part, $\zeta_{\text{P}k}$. All the common parts are then concatenated to form a combined common data stream, $\zeta_\text{C}=[\zeta_{\text{C}1},\ldots,\zeta_{\text{C}K}]$. This is then encoded into a real-valued common stream, $\Psi_\text{C}$, by using a code-book available at all the users \cite{9831449}. On the other hand, each private part is encoded into a real-valued private stream, $\Psi_{\text{P}k}$. Let $\alpha\in[0,1]$ denote the fraction of the total transmit power allocated to the communication signal, while $1-\alpha$ denotes the fraction allocated to the sensing signal. Furthermore, within the communication signal, the power allocation coefficients for the common stream and the $K$ private streams are denoted by $\beta_{\text{C}}$ and $\beta_{\text{P}k}$, respectively, such that $\beta_\text{C}+\sum_{k=1}^K\beta_{\text{P}k}=1$. Following the procedure outlined in \cite{9831449}, the private streams are combined and mapped onto the quadrature component, while the common stream is mapped onto the in-phase component. Thus, the transmit signal by the BS for the proposed PASS-QRSMA-ISAC framework can be given as:
\begin{equation}
\begin{split}
    \Psi(t)=\underbrace{\sqrt{\alpha}\Bigg(\underbrace{\sqrt{\beta_\text{C}}\Psi_\text{C}(t)}_{\text{In-Phase}}+j\underbrace{\sum_{k=1}^K\sqrt{\beta_{\text{P}k}}\Psi_{\text{P}k}(t)}_{\text{Quadrature}}\Bigg)}_{\text{Communication Signal}}+\underbrace{\sqrt{(1-\alpha)}\Psi_S}_{\text{Sensing Signal}}.
\end{split}
\end{equation}
This composite signal propagates along the waveguide, and different portions of it are radiated into the wireless medium through the $L$ PAs deployed on the waveguide in Room 1. It should be noted that, in this work, the total transmit power is assumed to be equally distributed among the $L$ PAs \cite{10945421}.  Since the transmitted signal supports both communication and sensing functionalities, this sub-section is further divided into two parts to discuss each task separately. 
\subsubsection{\textbf{Communication Task}} 
For the communication task, the CUs are distributed across the two rooms as described earlier and therefore experience the corresponding LoS or NLoS propagation conditions. Based on this, the received signal at $U_k$ in any room, can be expressed as:
\begin{equation}
    \EuScript{Y}_k(t)=\sqrt{\frac{\EuScript{P}_t}{L}}h_{\Sigma,k}\Psi(t)+z_k(t),
\end{equation}
where $z_k(t)\sim\mathbb{CN}\{0,\sigma^2\}$ represents the additive white Gaussian noise (AWGN) at $U_k$. At $U_k$, the received signal is first multiplied with $\frac{h^*_{\Sigma,k}}{\lvert h_{\Sigma,k}\rvert}$ to obtain an intermediate signal which can be given as:
\begin{equation}
    \widehat{\EuScript{Y}}_k(t)=\sqrt{\frac{\EuScript{P}_t}{L}}\lvert h_{\Sigma,k}\rvert\Psi(t)+\widehat{z}_k(t),
\end{equation}
where $\widehat{z}_k(t)\sim\mathbb{CN}\{0,\sigma^2\}$ is the modified noise term and has the same statistics as $z_k(t)$. The phase compensated signal $\widehat{\EuScript{Y}}_k(t)$ is then separated into its in-phase and quadrature components. The in-phase component is used to recover the common stream, whereas the quadrature component contains the superposition of the private streams. Accordingly, these two components can be written as:
\begin{equation}
\label{U2_Common}
\begin{split}
    \Re\left\{\widehat{\EuScript{Y}}_k(t)\right\}&=\sqrt{\frac{\alpha\beta_{\text{C}}\EuScript{P}_t}{L}}\lvert h_{\Sigma,k}\rvert\Psi_\text{C}(t)\\&\hspace{8mm}+\sqrt{\frac{(1-\alpha)\EuScript{P}_t}{L}}\lvert h_{\Sigma,k}\rvert\Re\left\{\Psi_S\right\}+\Re\left\{\widehat{z}_k(t)\right\},
\end{split}
\end{equation}
\begin{equation}
\label{U2_Private}
\begin{split}
    \Im\left\{\widehat{\EuScript{Y}}_k(t)\right\}&=\sqrt{\frac{\alpha\EuScript{P}_t}{L}}\lvert h_{\Sigma,k}\rvert\sum_{i=1}^K\sqrt{\beta_{\text{P}i}}\Psi_{\text{P}i}(t)\\&\hspace{8mm}+\sqrt{\frac{(1-\alpha)\EuScript{P}_t}{L}}\lvert h_{\Sigma,k}\rvert\Im\left\{\Psi_S\right\}+\Im\left\{\widehat{z}_k(t)\right\},
\end{split}
\end{equation}
where $\Re\left\{\widehat{z_k}(t)\right\}$ and $\Im\left\{\widehat{z_k}(t)\right\}\sim\mathbb{N}\left\{0,\frac{\sigma^2}{2}\right\}$. Thus, using \eqref{U2_Common} and \eqref{U2_Private}, the corresponding SNR/ SINR of the common and the private streams can be respectively given as:
\begin{equation}
    \label{Common_SINR}
    \Lambda_{k,Com,Pin}=\frac{\frac{\alpha\beta_{\text{C}}\EuScript{P}_t}{L}\lvert h_{\Sigma,k}\rvert^2}{\frac{1}{2}\left(\frac{(1-\alpha)\EuScript{P}_t}{L}\lvert h_{\Sigma,k}\rvert^2+\sigma^2\right)},
\end{equation}
\begin{equation}
\label{Private_SINR}
    \Lambda_{k,Pvt,Pin}=\frac{\frac{\alpha\EuScript{P}_t\beta_{\text{P}k}}{L}\lvert h_{\Sigma,k}\rvert^2}{\frac{\alpha\EuScript{P}_t}{L}\lvert h_{\Sigma,k}\rvert^2\sum_{\substack{i=1 \\i\neq k}}^K\beta_{\text{P}i}+\frac{1}{2}\left(\frac{(1-\alpha)\EuScript{P}_t}{L}\lvert h_{\Sigma,k}\rvert^2+\sigma^2\right)}.
\end{equation}
\paragraph{\textbf{Rate Calculation}} Using \eqref{Common_SINR} and \eqref{Private_SINR}, the corresponding instantaneous rate of the common and private streams of any $U_k$ can be respectively given as:
\begin{equation}
\label{instantaneous_rate}
    \EuScript{R}_{\text{C}k}=\frac{1}{2}\log_2\left(1+\Lambda_{k,Com,Pin}\right); \quad \EuScript{R}_{\text{P}k}=\frac{1}{2}\log_2\left(1+\Lambda_{k,Pvt,Pin}\right),
\end{equation}
where the factor $\frac{1}{2}$ is multiplied to account for the use of real-valued symbols for the common and private streams. Since a real symbol carries half the DoF of a complex symbol, the corresponding scaling is required \cite{9831449}. Now, since the common stream is to be decoded at both the users, the achievable rate of the common stream can be given as:
\begin{equation}
    \EuScript{R}_{\text{C}}=\min(\EuScript{R}_{\text{C}1},\EuScript{R}_{\text{C}2},\ldots,\EuScript{R}_{\text{C}K}),
\end{equation}
where $r_k$ is the rate allocated for $U_k$ such that $\sum_{k=1}^K r_k\leq\EuScript{R}_{\text{C}}$. Thus, the achievable rate of $U_k$ can be given as $\EuScript{R}_k=r_k+\EuScript{R}_{\text{P}k}$. Subsequently, the sum rate of the system can be given as:
\begin{equation}
\label{SumRate}
    \EuScript{R}_{\Sigma}=\sum\nolimits_{k=1}^K r_k+\EuScript{R}_{\text{P}k}=\EuScript{R}_{\text{C}}+\sum\nolimits_{k=1}^K\EuScript{R}_{\text{P}k}.
\end{equation}
\subsubsection{\textbf{Sensing Task}}
The sensing signal is transmitted through the $L$ PAs simultaneously with the communication signal, and its reflected echo from the target is subsequently received at the sensing receiver. It can be done so by activating the closest PA to the ST which has been denoted using $\widehat{P_l}$. Accordingly, the received sensing echo signal, which is used for target range estimation, can be expressed as follows:
\begin{equation}
\begin{split}
    \EuScript{Y}_{S}(t)&=\sqrt{\frac{\EuScript{P}_t}{L}}h_{sen}\Psi(t-T_S)+z_S(t)\\&=\sqrt{\frac{\EuScript{P}_t\xi_S}{L}}h_{\Sigma,S}\widehat{h_{\Sigma,S}}\Psi(t-T_S)+z_S(t),
\end{split}    
\end{equation}
where $T_S$ is the round trip propagation delay, and $z_S(t)\sim\mathbb{CN}\{0,\sigma_S^2\}$. Practically, the signals radiated from different PAs may experience slightly different round-trip propagation delays due to their distinct propagation paths. However, a common round-trip delay $T_S$ is adopted following the common-delay assumption in \cite{11274872}. For the considered single-PA echo reception, this approximation is valid when the maximum PA deployment length satisfies $D\leq \frac{c}{\EuScript{B}}$, where $\EuScript{B}$ is the bandwidth. Here, $\Psi(t-T_S)$ contains both the communication and sensing components. Since, the communication component of $\Psi(t-T_S)$ is known at the BS, it can be effectively removed using analog and digital interference cancellation techniques. However, owing to hardware constraints, this removal might not be perfect. Thus, the SINR of the received echo sensing signal can be given as:
\begin{equation}
    \label{Sensing_SNR}
    \Lambda_{Sensing,Pin}=\frac{\frac{\EuScript{P}_t\xi_S}{L}\lvert h_{\Sigma,S}\rvert^2\left\lvert \widehat{h_{\Sigma,S}}\right\rvert^2(1-\alpha)}{\sigma_S^2+\Delta_S\frac{\EuScript{P}_t\xi_S\alpha}{L}\lvert h_{\Sigma,S}\rvert^2\left\lvert \widehat{h_{\Sigma,S}}\right\rvert^2},
\end{equation}
where $\Delta_S\in[0,1]$ denotes the residual interference coefficient caused by imperfect cancellation of the communication component during sensing-signal recovery. In this work, we consider hardware induced imperfect cancellation.
\section{Two-User Single-PA Performance Analysis}
\label{singlePA_analysis}
In this section, we analyse the outage probability and ergodic sum-rate performance of the considered system. To obtain tractable baseline analytical insights, we first focus on the two-user single-PA case under an ideal waveguide scenario, which is consistent with existing two-user single-PA PASS-RSMA-ISAC analytical studies where ideal waveguide propagation is considered to simplify the analysis \cite{ShivSir}. Accordingly, we consider $L=1$, $\Phi_P\triangleq \Phi_{P_1}$, $K=2$, $M=1$, and $\kappa=0$, with the PA aligned with $U_S$ along the $x$-axis such that $\Phi_P=[x_S,0,d]$ to shorten the propagation distance and enhance the echo signal.
\subsection{\textbf{Outage Analysis}}
In this sub-section, we provide the user COP, system COP and SOP analysis, respectively for the proposed single PA framework. 
\subsubsection{\textbf{COP Analysis}}\label{COP_Section} In a Q-RSMA-based communication system, an outage event occurs when either the common stream or the corresponding private stream cannot be successfully decoded. Since the channel statistics differ for users located in different rooms, we first derive the outage probability on a user-wise basis which is then followed by the system outage probability.
\begin{figure*}[!t]
    \centering
    \begin{equation}
    \label{CDF_U1}
    \begin{split}
    \EuScript{F}_{\EuScript{X}_1}(x_1)
    =
    \begin{cases}
        0, & x_1<d^2,\\
        \frac{\pi R^2}{D^2}-\frac{4R^3}{3D^3}, & d^2\leq x_1<d^2+\frac{D^2}{4},\\
        \frac{1}{D^2}\left[\frac{D^2}{12}+D\Omega-2R^2\text{arcsin}\left(\frac{\Omega}{R}-R^2+\pi R^2\right)\right], & d^2+\frac{D^2}{4}\leq x_1<d^2+D^2,\\
        \frac{\Omega}{D}-\frac{\Omega^2}{D^2}+\frac{2R^2}{D^2}\Big[\text{arcsin}\left(\frac{D}{R}\right)-\text{arcsin}\left(\frac{\Omega}{R}\right) \Big]+\frac{2U}{D}+\frac{4U^3}{3D^3}-\frac{1}{6}, & d^2+D^2\leq x_1\leq d^2+\frac{5D^2}{4},\\
        1, & x_1>d^2+\frac{5D^2}{4}.
    \end{cases}
    \end{split}
    \end{equation}
    \hrule
    \vspace{-1.25em}
\end{figure*}
\begin{itemize}
    \item \textbf{User 1:} The COP of $U_1$ can be expressed as:
    \begin{equation}
    \begin{split}
        \text{P}_{O1}&=1-\text{Pr}\left\{\Lambda_{1,Com,Pin}>\tau_{C1}, \Lambda_{1,Pvt,Pin}>\tau_{P1}\right\}\\
        &=1-\text{Pr}\left\{\EuScript{X}_1^{-1}>\delta_{1}\right\}=1-\EuScript{F}_{\EuScript{X}_1}\left(\frac{1}{\delta_{1}}\right),
    \end{split}
    \end{equation}
    where $\EuScript{X}_1=\lVert \Phi_\text{P}-\Phi_1\rVert^2$ and $\delta_{1}$=$\max\left(\delta_{C1},\delta_{P1}\right)$ with 
    \begin{subequations}
    \label{Threshold_U1}
    \begin{align}
        \delta_{C1}(\tau_{C1})&=\frac{\tau_{C1}\sigma^2}{\EuScript{P}_t\xi^2\left(2\alpha\beta_\text{C}-\tau_{C1}(1-\alpha)\right)},\label{Threshold_Common_U1}\\
        \delta_{P1}(\tau_{P1})&=\frac{\tau_{P1}\sigma^2}{\EuScript{P}_t\xi^2\left(2\alpha\beta_{\text{P}1}-\tau_{P1}(2\alpha\beta_{\text{P}2}+1-\alpha)\right)}\label{Threshold_Pvt_U1},
    \end{align}    
    \end{subequations}
    where $\tau_{C1}=2^{2\mathbb{R}_{C1,Th}}-1$ and $\tau_{P1}=2^{2\mathbb{R}_{P1,Th}}-1$. Here, $\mathbb{R}_{C1,Th}$ and $\mathbb{R}_{P1,Th}$ are the target data rates of the common and private streams of $U_1$, respectively. Thus, the CDF of $\EuScript{X}_1$ can be expressed from \cite{ShivSir} as shown in \eqref{CDF_U1} where $R$=$\sqrt{x_1-d^2}$, $\Omega$=$\sqrt{\max\left(0, x_1-d^2-\frac{D^2}{4}\right)}$, and $U$=$\sqrt{\max(0, x_1-d^2-D^2)}$.
    \item \textbf{User 2:} The COP of $U_2$ can be expressed as:
    \begin{figure*}[t]
    \centering
    \begin{equation}
    \label{PDF_U2}
    \begin{split}
    f_{\EuScript{X}_2}(x_2)
    =
    \begin{cases}
        0, & x_2<d^2,\\
        \frac{\sqrt{x_2-d^2}}{D^3}, & d^2\leq x_2<d^2+\frac{D^2}{4},\\
        \frac{1}{2D^2}, & d^2+\frac{D^2}{4}\leq x_2<d^2+D^2,\\
        \frac{1}{2D^2}-\frac{2\sqrt{x_2-d^2-D^2}}{D^3}+\frac{2}{D^2}\left[\arctan\left(\frac{\sqrt{x_2-d^2-D^2}}{D}\right)-\frac{\sqrt{x_2-d^2-D^2}}{D}+\frac{D\sqrt{x_2-d^2-D^2}}{\sqrt{x^2-d^2}}\right], & d^2+D^2\leq x_2<d^2+\frac{5D^2}{4}, \\
        \frac{2\pi-1}{2D^2}-\frac{2}{D^2}\arctan\left(\frac{\sqrt{4(x_2-d^2)-D^2}}{D}\right), & d^2+\frac{5D^2}{4}\leq x_2<d^2+4D^2,\\
        \frac{2}{D\sqrt{4\left(x_2-d^2\right)-D^2}}-\frac{1}{2D^2}+\frac{\sqrt{x_2-d^2-4D^2}}{D^3}+\frac{2}{D^2}\Bigg[\arctan\left(\frac{D}{\sqrt{4(x_2-d^2)-D^2}}\right) \\\hspace{50mm}-\arctan\left(\frac{\sqrt{x_2-d^2-4D^2}}{2D}\right)-\frac{D\sqrt{4\left(x_2-d^2\right)-D^2}}{4\sqrt{x_2-d^2}}\Bigg], & d^2+4D^2\leq x_2<d^2+\frac{17D^2}{4},\\
        0, & x_2\geq d^2+\frac{17D^2}{4}.
    \end{cases}
    \end{split}
    \end{equation}
    \hrule
    \end{figure*}
    \begin{equation}
    \label{Outage_U2}
    \begin{split}
        \text{P}_{O2}&=1-\text{Pr}\left\{\Lambda_{2,Com,Pin}>\tau_{C2}, \Lambda_{2,Pvt,Pin}>\tau_{P2}\right\}\\
        &=1-\text{Pr}\left\{\EuScript{X}_2^{\frac{-\nu}{2}}\lvert g_2 \rvert^2>\delta_{2}\right\},
    \end{split}
    \end{equation}
    where $\EuScript{X}_2=\lVert \Phi_\text{P}-\Phi_2\rVert^2$ and $\delta_2$=$\max\left(\delta_{C2},\delta_{P2}\right)$ with
    \begin{subequations}
    \begin{align}
        \delta_{C2}(\tau_{C2})&=\frac{\tau_{\text{C}2}\sigma^2}{\EuScript{P}_t(2\alpha\beta_\text{C}-\tau_{\text{C}2}(1-\alpha))}\label{Threshold_Common_U2},\\
        \delta_{P2}(\tau_{P2})&=\frac{\tau_{P2}\sigma^2}{\EuScript{P}_t(2\alpha\beta_{\text{P}2}-\tau_{\text{P}2}(2\alpha\beta_{\text{P}1}+1-\alpha))}\label{Threshold_Pvt_U2},
    \end{align}    
    \end{subequations}
    where $\tau_{C2}=2^{2\mathbb{R}_{C2,Th}}-1$ and $\tau_{P2}=2^{2\mathbb{R}_{P2,Th}}-1$. Here, $\mathbb{R}_{C2,Th}$ and $\mathbb{R}_{P2,Th}$ are the target data rates of the common and private streams of $U_2$, respectively and the PDF of $\EuScript{X}_2$ can be expressed from \cite{cheng2025performance} and is shown in \eqref{PDF_U2}. Thus, on solving \eqref{Outage_U2}, the COP of $U_2$ can be approximated as:
    \begin{equation}
    \label{Outage_U2_Ana}
        \text{P}_{O2}=1-\sum\nolimits_{n=1}^5\mathbb{P}_{2n}(\delta_2),
    \end{equation}
    where $\mathbb{P}_{21}$, $\mathbb{P}_{22}$, $\mathbb{P}_{23}$, $\mathbb{P}_{24}$, and $\mathbb{P}_{25}$ can be found in \eqref{P21}-\eqref{P25}, respectively with $t_m=\cos\left(\frac{2m-1}{2\EuScript{M}}\pi\right)$ and $\EuScript{M}$ is the number of Chebyshev quadrature points. The derivation of \eqref{Outage_U2_Ana} from \eqref{Outage_U2} is provided in Lemma \ref{Lemma_proof}. It should be noted that the above COP expressions are valid for $2\alpha\beta_\text{C}-\tau_{\text{C}k}(1-\alpha)>0$ and $2\alpha\beta_{\text{P}k}-\tau_{\text{P}k}(2\alpha\sum_{i=1,i\neq k}^2\beta_{\text{P}i}+1-\alpha)>0$. Otherwise, the corresponding decoding event becomes infeasible and $\text{P}_{Ok}=1$.
    \begin{lemma}
    \label{Lemma_proof}
    The COP of $U_2$ as shown in \eqref{Outage_U2}, can be approximated as follows:
    \begin{equation*}
        \text{P}_{O2}=1-\{\mathbb{P}_{21}+\mathbb{P}_{22}+\mathbb{P}_{23}+\mathbb{P}_{24}+\mathbb{P}_{25}\},
    \end{equation*}
    where we condition on $\EuScript{X}_2^{\frac{-\nu}{2}}$.
        \begin{figure*}[t]
    \centering
    \begin{subequations}
    \label{USER2_OutageEquations}
    \begin{align}
    \mathbb{P}_{21}(\delta_2)&=\frac{\pi}{8\EuScript{M}}\sum_{m=1}^\EuScript{M}\sqrt{1-t_m^2}\left(\frac{\sqrt{1+t_m}}{2\sqrt{2}}\right)e^{-\delta_2\left(\frac{D^2t_m+8d^2+D^2}{8}\right)^{\frac{\nu}{2}}}, \label{P21} \\
    \mathbb{P}_{22}(\delta_2)&=\frac{3\pi}{16\EuScript{M}}\sum_{m=1}^\EuScript{M}\sqrt{1-t_m^2} e^{-\delta_2\left(\frac{3D^2t_m+8d^2+5D^2}{8}\right)^{\frac{\nu}{2}}}, \label{P22} \\
    \mathbb{P}_{23}(\delta_2)&=\frac{\pi}{8\EuScript{M}}\sum_{m=1}^\EuScript{M}\sqrt{1-t_m^2}\left[\frac{1}{2}-\sqrt{\frac{1+t_m}{2}}+2\text{arctan}\left(\frac{\sqrt{1+t_m}}{2\sqrt{2}}\right)\right] e^{-\delta_2\left(\frac{D^2t_m+8d^2+9D^2}{8}\right)^{\frac{\nu}{2}}},\label{P23} \\
    \mathbb{P}_{24}(\delta_2)&=\frac{11\pi}{8\EuScript{M}}\sum_{m=1}^\EuScript{M}\sqrt{1-t_m^2}\left[\frac{2\pi-1}{2}-2\text{arctan}\left(\sqrt{\frac{11t_m+19}{2}}\right)\right] e^{-\delta_2\left(\frac{11D^2t_m+8d^2+21D^2}{8}\right)^{\frac{\nu}{2}}}, \label{P24} \\
    \mathbb{P}_{25}(\delta_2)&=\frac{\pi}{8\EuScript{M}}\sum_{m=1}^\EuScript{M}\sqrt{1-t_m^2}\left[-\frac{1}{2}+\frac{\sqrt{1+t_m}}{2\sqrt{2}}+2\left(\text{arctan}\left(\sqrt{\frac{2}{t_m+31}}\right)-\text{arctan}\left(\sqrt{\frac{1+t_m}{4\sqrt{2}}}\right)\right)\right] e^{-\delta_2\left(\frac{D^2t_m+8d^2+33D^2}{8}\right)^{\frac{\nu}{2}}}.\label{P25} 
    \end{align}
    \end{subequations}
    \hrule
     \vspace{-1.25em}
    \end{figure*}
    \begin{proof}
        Considering $\lvert g_2\rvert^2$ to be exponentially distributed with parameter 1 and conditioning on $\EuScript{X}_2=x_2$, \eqref{Outage_U2} can be expressed as:
        \vspace{-0.0em}
        \begin{equation}
        \label{ANA_U2_Proof}
        \begin{split}
        \text{P}_{O2}&=1-\text{Pr}\left\{\EuScript{X}_2^{\frac{-\nu}{2}}\lvert g_2 \rvert^2>\delta_{2}\right\}=\text{Pr}\left\{\lvert g_2 \rvert^2\leq\delta_{2}\EuScript{X}_2^{\frac{\nu}{2}}\right\}\\
        &=\int_0^\infty\left(1-e^{-\delta_2x_2^{\frac{\nu}{2}}}\right)f_{\EuScript{X}_2}(x_2)dx_2\\&=1-\int_0^\infty e^{-\delta_2x_2^{\frac{\nu}{2}}}f_{\EuScript{X}_2}(x_2)dx_2.
        \end{split}
        \end{equation}
        Now, using \eqref{PDF_U2}, \eqref{ANA_U2_Proof} can be solved as:
        \begin{equation}
            \begin{split}
                &\text{P}_{O2}=1-\Bigg\{\underbrace{\int_{d^2}^{d^2+\frac{D^2}{4}}e^{-\delta_2x_2^{\frac{\nu}{2}}}f_{\EuScript{X}_2}(x_2)dx_2}_{\mathbb{P}_{21}}+\\
                &\underbrace{\int_{d^2+\frac{D^2}{4}}^{d^2+D^2}e^{-\delta_2x_2^{\frac{\nu}{2}}}f_{\EuScript{X}_2}(x_2)dx_2}_{\mathbb{P}_{22}}+\underbrace{\int_{d^2+D^2}^{d^2+\frac{5D^2}{4}}e^{-\delta_2x_2^{\frac{\nu}{2}}}f_{\EuScript{X}_2}(x_2)dx_2}_{\mathbb{P}_{23}}+\\
                &\underbrace{\int_{d^2+\frac{5D^2}{4}}^{d^2+4D^2}e^{-\delta_2x_2^{\frac{\nu}{2}}}f_{\EuScript{X}_2}(x_2)dx_2}_{\mathbb{P}_{24}}+\underbrace{\int_{d^2+4D^2}^{d^2+\frac{17D^2}{4}}e^{-\delta_2x_2^{\frac{\nu}{2}}}f_{\EuScript{X}_2}(x_2)dx_2}_{\mathbb{P}_{25}}\Bigg\}.
            \end{split}
        \end{equation}
        To solve the above piecewise integrals, the Gaussian Chebyshev Quadrature method of the first kind (GCQ-I) is employed \cite{ShivSir}. To do so, a variable transformation needs to be performed to map the range onto the interval [-1,1]. Now, to approximate a standard integral of the form $\int_{-1}^1b(t)dt$ with GCQ-I, the following can be used:
        \begin{equation}
            \int_{-1}^1b(t)dt\approx\sum\nolimits_{m=1}^\EuScript{M} \frac{\pi}{\EuScript{M}}b(t_m)\sqrt{1-t_m^2}.
        \end{equation}
        Considering $x_2=\frac{D^2t_m+8d^2+D^2}{8}$, we have
        \begin{equation}
        \begin{split}
            \mathbb{P}_{21}&=\int_{-1}^1 \frac{D^2}{8}e^{-\delta_2\left(\frac{D^2t_m+8d^2+D^2}{8}\right)^{\frac{\nu}{2}}}\frac{\sqrt{1+t_m}}{2\sqrt{2}D^2}dt_m\\
            &=\frac{\pi}{8\EuScript{M}}\sum_{m=1}^\EuScript{M}\sqrt{1-t_m^2}\left(\frac{\sqrt{1+t_m}}{2\sqrt{2}}\right)e^{-\delta_2\left(\frac{D^2t_m+8d^2+D^2}{8}\right)^{\frac{\nu}{2}}}.
        \end{split}    
        \end{equation}
        Similarly, $\mathbb{P}_{22}$, $\mathbb{P}_{23}$, $\mathbb{P}_{24}$, and $\mathbb{P}_{25}$ can be derived.
    \end{proof}
    \end{lemma}
    \item \textbf{System Outage:} The system is said to be in outage if either of the users are in an outage. Thus, using the inclusion-exclusion principle, and considering the events to be independent, the system COP can be given as:
    \begin{equation}
    \label{system_outage_singlaPA}
        \text{P}_{Sys}=\text{P}_{O1}+\text{P}_{O2}-\text{P}_{O1}\text{P}_{O2}=1-(1-\text{P}_{O1})(1-\text{P}_{O2}).
    \end{equation}
\end{itemize}
\subsubsection{\textbf{SOP Analysis}}
\label{SOP_Analysis_SinglePA}
As discussed earlier, the COP is a fundamental information-theoretic metric used to quantify communication reliability. However, it is not directly applicable to wireless sensing tasks, since sensing performance is governed by estimation accuracy rather than achievable rate. To enable a unified performance evaluation framework for the proposed PASS-QRSMA-ISAC system, we adopt a modified outage definition based on the MSE, in line with the methodology of \cite{ShivSir, Heath, 10542219}. Specifically, the sensing MSE in this work corresponds to the target-range estimation error, where the received sensing echo is used to estimate the round-trip propagation delay $T_S$, which is directly related to the target range. A detailed Fisher-information-based derivation for the corresponding CRLB in PASS-assisted ISAC range estimation can be found in \cite{ShivSir}. Accordingly, the MSE-based SOP is defined as the probability that the sensing MSE exceeds a prescribed threshold which is given as $\text{P}_{OS}=\Pr(\rho_S>\rho_{Th})$, where $\rho_S$ denotes the sensing MSE and $\rho_{Th}$ is the target MSE threshold. For a sensing signal with a flat spectral density, the CRLB for the range-estimation MSE can be expressed as $\rho_S\geq\rho_S^{CRLB}=\frac{c^2}{k_S\Lambda_{Sensing,Pin}(\varpi\EuScript{B})^2},$ where $k_S=\frac{64\pi^2}{12}$, and $\varpi$ is the spectrum allocation factor for the sensing task. Therefore, the CRLB-based MSE outage condition can be equivalently mapped to the sensing-SINR threshold condition $\Lambda_{Sensing,Pin}\leq \tau_S$, where $\tau_S=\frac{c^2}{k_S\left(\varpi\EuScript{B}\right)^2\rho_{Th}}$. Thus, the SOP can be evaluated using the distribution of $\Lambda_{Sensing,Pin}$ as follows:
\begin{equation}
\label{SOP_Normal}
\begin{split}
    \text{P}_{OS}&=\Pr(\Lambda_{Sensing,Pin}\leq \tau_{S})\\
    &=\Pr\left(\frac{\xi_S(1-\alpha)\EuScript{P}_t\lVert \Phi_\text{P}-\Phi_S \rVert^{-4}}{\sigma_S^2 + \Delta_{S} \xi_S\alpha\EuScript{P}_t\lVert \Phi_\text{P}-\Phi_S \rVert^{-4}}\leq \tau_{S}\right)\\
    &=\Pr\left(\lVert \Phi_\text{P}-\Phi_S \rVert^{-4}\leq\frac{\tau_S\sigma_S^2}{\xi_S\EuScript{P}_t\left[(1-\alpha)-\alpha\tau_S\Delta_S\right]}\right)\\
    &=1-\Pr\left(\EuScript{X}_S\leq\sqrt{\frac{\xi_S\EuScript{P}_t\left[(1-\alpha)-\alpha\tau_S\Delta_S\right]}{\tau_S\sigma_S^2}}\right)\\
    &=1-\EuScript{F}_{\EuScript{X}_S}\left(\sqrt{\frac{\xi_S\EuScript{P}_t\left[(1-\alpha)-\alpha\tau_S\Delta_S\right]}{\tau_S\sigma_S^2}}\right),
\end{split}
\end{equation}
where $\EuScript{X}_S=\lVert \Phi_\text{P}-\Phi_S \rVert^{2}$, $\tau_S=\frac{c^2}{k_S\rho_{Th}(\varpi\EuScript{B})^2}$ is the sensing SINR threshold expression, and $\EuScript{F}_{\EuScript{X}_S}$ is the CDF corresponding to $\EuScript{X}_S$ which can be given as:
\begin{equation}
\label{SOP_CDF}
    \EuScript{F}_{\EuScript{X}_S}(x_S)=
    \begin{cases}
        0, & x_S<d^2,\\
        \frac{2\sqrt{x_S-d^2}}{D}, & d^2\leq x_S<d^2+\frac{D^2}{4},\\
        1, & x_S\geq d^2+\frac{D^2}{4}.
    \end{cases}
\end{equation}
It should be noted that the SOP expression is valid when $(1-\alpha)-\alpha\tau_S\Delta_S>0$. If the condition is not satisfied, then $\text{P}_{OS}=1$.
\subsection{\textbf{Ergodic Sum Rate Analysis}}
The ergodic sum-rate represents the average achievable sum-rate over fading/channel realisations. Accordingly, for the proposed framework, it is given as:
\begin{equation}
\label{SumRate_Ana}
\begin{split}
    \EuScript{R}_{\Sigma,Ana}&=\mathbb{E}\left\{\min(\EuScript{R}_{\text{C}1},\EuScript{R}_{\text{C}2})+\sum_{i=1}^2\EuScript{R}_{\text{P}i}\right\}\\
    &=\underbrace{\mathbb{E}\left\{\min(\EuScript{R}_{\text{C}1},\EuScript{R}_{\text{C}2})\right\}}_{\text{ACRA}}+\sum_{i=1}^2\underbrace{\mathbb{E}\left\{\EuScript{R}_{\text{P}i}\right\}}_{\text{APRA}},
\end{split}
\end{equation}
where ACRA stands for achievable common rate analysis and APRA stands for achievable private rate analysis.
\begin{itemize}
    \item \textbf{Achievable Common Rate Analysis:} Now, we derive the achievable common rate average by taking the ergodic expectation of the instantaneous common rate, which is limited by the weaker common decoding capability among the two users. Thus, the ACRA can be given as:  \begin{equation}
    \label{ACRA}
    \begin{split}
        \mathbb{E}\left\{\EuScript{R}_\text{C}\right\}&=\frac{1}{2}\mathbb{E}\left\{\log_2(1+\min(\Lambda_{1,Com,Pin}, \Lambda_{2,Com,Pin}))\right\}\\
        &=\frac{1}{2\ln2}\int_0^\infty \frac{\text{Pr}(\Lambda_{1,Com,Pin}>x, \Lambda_{2,Com,Pin}>x)}{1+x}dx\\
        &=\frac{1}{2\ln2}\int_0^\infty \frac{F_{\Lambda_{C_1}}(x)+F_{\Lambda_{C_2}}(x)-1+F_{\Lambda_{C_1},\Lambda_{C_2}}(x)}{1+x}dx,
    \end{split}
    \end{equation}
    where $F_{\Lambda_{C_1}}$ and $F_{\Lambda_{C_2}}$ denote the complementary cumulative distribution functions (CCDFs) of $\Lambda_{1,Com,Pin}$ and $\Lambda_{2,Com,Pin}$, respectively, and $F_{\Lambda_{C_1},\Lambda_{C_2}}(x)$ denotes their joint CDF which can be expressed as shown below:
    \begin{equation}
    \label{FC1}
        F_{\Lambda_{C_1}}(x)\triangleq\text{Pr}(\Lambda_{1,Com,Pin}>x)=\EuScript{F}_{\EuScript{X}_1}\left(\frac{1}{\delta_{C1}(x)}\right),
    \end{equation}
    \begin{equation}
    \label{FC2}
        \begin{split}
            F_{\Lambda_{C_2}}(x)&\triangleq \text{Pr}(\Lambda_{2,Com,Pin}>x)=\sum_{n=1}^5\mathbb{P}_{2n}(\delta_{C2}(x)),
        \end{split}
    \end{equation}
    \begin{equation}
        F_{\Lambda_{C_1},\Lambda_{C_2}}(x)=\EuScript{C}\left(1-F_{\Lambda_{C_1}}(x), 1-F_{\Lambda_{C_2}}(x);\theta_2\right),
    \end{equation}
    where $\EuScript{C}(u,v)$ denotes a bivariate copula function, which characterizes the joint cumulative distribution function (CDF) by linking the marginal CDFs of the involved random variables \cite{11367373}, $\delta_{C1}(x)$ can be obtained from \eqref{Threshold_Common_U1}, $\delta_{C2}(x)$ can be obtained from \eqref{Threshold_Common_U2} and $\theta_2\in[-1,1]$. Now, as explained earlier in Section \ref{COP_Section}, the dependency is observed to be negligible which has been verified through the close agreement between the simulated and analytical results under the independence assumption in Section \ref{results}. Thus, \eqref{ACRA} can also be expressed as follows:
    \begin{equation}
    \label{ACRA_Independent}
    \begin{split}
        \mathbb{E}\left\{\EuScript{R}_\text{C}\right\}&=\frac{1}{2\ln2}\int_0^\infty \frac{F_{\Lambda_{C_1}}(x)F_{\Lambda_{C_2}}(x)}{1+x}dx.
    \end{split}
    \end{equation}
    \begin{lemma} 
    The ACRA term shown in \eqref{ACRA}, can be approximated as follows:
    \begin{equation*}
        \mathbb{E}\left\{\EuScript{R}_\text{C}\right\}=\frac{1}{2\ln2}\int_0^\infty \frac{F_{\Lambda_{C_1}}(x)F_{\Lambda_{C_2}}(x)}{1+x}dx,
    \end{equation*}
    where the variables $\Lambda_{1,Com,Pin}$ and $\Lambda_{2,Com,Pin}$ are considered to be independent.
    \begin{proof}
        From \eqref{FC1} and \eqref{FC2}, we can also have the equations as shown below:
        \begin{subequations}
        \begin{align}
            1-F_{\Lambda_{C_1}}(x)&\triangleq\text{Pr}(\Lambda_{1,Com,Pin}\leq x),\\ 
            1-F_{\Lambda_{C_2}}(x)&\triangleq \text{Pr}(\Lambda_{2,Com,Pin}\leq x)
        \end{align}
        \end{subequations}
        Now, the inclusion-exclusion principle can be applied on $\text{Pr}(\{\Lambda_{1,Com,Pin}\leq x\}\cup\{\Lambda_{2,Com,Pin}\leq x\})$ to obtain:
        \begin{equation}
        \label{IEP}
        \begin{split}
            &\text{Pr}\left(\{\Lambda_{1,Com,Pin}\leq x\}\cup\{\Lambda_{2,Com,Pin}\leq x\}\right)\\
            &=\text{Pr}\left(\Lambda_{1,Com,Pin}\leq x\right)+\text{Pr}\left(\Lambda_{2,Com,Pin}\leq x\right)-\\&\hspace{20mm}\text{Pr}\left(\{\Lambda_{1,Com,Pin}\leq x\}\cap\{\Lambda_{2,Com,Pin}\leq x\}\right)\\
            &=(1-F_{\Lambda_{C_1}}(x)) + (1-F_{\Lambda_{C_2}}(x))-F_{\Lambda_{C_1},\Lambda_{C_2}}(x).
        \end{split}
        \end{equation}
        Thus, applying complement rule to \eqref{IEP}, we obtain $\text{Pr}(\Lambda_{1,Com,Pin}>x, \Lambda_{2,Com,Pin}>x)$ which can be given as:
        \begin{equation}
        \label{IEP_2}
        \begin{split}
            &\text{Pr}(\Lambda_{1,Com,Pin}>x, \Lambda_{2,Com,Pin}>x)\\&=1-\text{Pr}(\{\Lambda_{1,Com,Pin}\leq x\}\cup\{\Lambda_{2,Com,Pin}\leq x\})\\
            &=1-\left[(1-F_{\Lambda_{C_1}}(x)) + (1-F_{\Lambda_{C_2}}(x))-F_{\Lambda_{C_1},\Lambda_{C_2}}(x)\right]\\
            &=F_{\Lambda_{C_1}}(x) + F_{\Lambda_{C_2}}(x)-1+F_{\Lambda_{C_1},\Lambda_{C_2}}(x).
        \end{split}
        \end{equation}
        Now, considering $\Lambda_{1,Com,Pin}$ and $\Lambda_{2,Com,Pin}$ as independent, we obtain $F_{\Lambda_{C_1},\Lambda_{C_2}}(x)=(1-F_{\Lambda_{C_1}}(x))(1-F_{\Lambda_{C_2}}(x))$. Now, on substituting this in \eqref{IEP_2} and solving it, we obtain:
        \begin{equation}
            \text{Pr}(\Lambda_{1,Com,Pin}>x, \Lambda_{2,Com,Pin}>x)=F_{\Lambda_{C_1}}(x)F_{\Lambda_{C_2}}(x).
        \end{equation}
        This can be substituted in \eqref{ACRA} to obtain \eqref{ACRA_Independent}.
    \end{proof}
    \end{lemma}
    \item \textbf{Achievable Private Rate Analysis:} In contrast to the common rate, the private rate is determined by each user’s own decoding capability without being limited by the weaker user. Thus, the APRA of any user, $U_i$ can be given as:
    \begin{equation}
    \begin{split}
        \mathbb{E}\left\{\EuScript{R}_{\text{P}i}\right\}&=\frac{1}{2}\mathbb{E}\left\{\log_2\left(1+\Lambda_{i,Pvt,Pin}\right)\right\}\\
        &=\frac{1}{2\ln 2}\int_0^\infty \frac{\text{Pr}(\Lambda_{i,Pvt,Pin}>x)}{1+x}dx\\
        &=\begin{cases}
            \frac{1}{2\ln 2}\int_0^\infty \frac{\EuScript{F}_{\EuScript{X}_1}\left(\frac{1}{\delta_{P1}(x)}\right)}{1+x}dx, & i=1,\\
            \frac{1}{2\ln 2}\int_0^\infty \frac{\sum_{n=1}^5\mathbb{P}_{2n}(\delta_{P2}(x))}{1+x}dx, & i=2,\\
        \end{cases}
    \end{split}
    \end{equation}
    where $\delta_{P1}(x)$ and $\delta_{P2}(x)$ can be obtained from \eqref{Threshold_Pvt_U1} and \eqref{Threshold_Pvt_U2}, respectively.
\end{itemize}
Thus, using the above ACRA and APRA equations, \eqref{SumRate_Ana} can be given as:
\begin{equation}
\label{SumRate_Final}
    \begin{split}
    \EuScript{R}_{\Sigma,Ana}&=\frac{1}{2\ln2}\Bigg\{\int_0^\infty \frac{F_{\Lambda_{C_1}}(x)F_{\Lambda_{C_2}}(x)}{1+x}dx+\\&\hspace{4mm}\int_0^\infty \frac{\EuScript{F}_{\EuScript{X}_1}\left(\frac{1}{\delta_{P1}(x)}\right)}{1+x}dx+\int_0^\infty \frac{\sum_{n=1}^5\mathbb{P}_{2n}(\delta_{P2}(x))}{1+x}dx\Bigg\}.
    \end{split}
\end{equation}
Due to the intractable nature of \eqref{SumRate_Final}, the corresponding integrals are evaluated numerically using the trapezoidal rule.
\begin{remark}
The derived expressions for the ergodic rates are valid within the feasible SINR region where the denominators of the corresponding $\delta$-based terms obtained from \eqref{Threshold_Common_U1}, \eqref{Threshold_Common_U2}, \eqref{Threshold_Pvt_U1}, and \eqref{Threshold_Pvt_U2} remain positive. For values of $x$ that violate this condition, the events $\Lambda_{i,Com,\text{Pin}}>x$ and $\Lambda_{i,Pvt,\text{Pin}}>x$ become infeasible, and hence the associated SINR complementary cumulative distribution function (CCDF) terms evaluate to zero.
\end{remark}
\section{Multi-User Multi-PA Performance Analysis}
\label{multiplePA_analysis}
In this section, we extend the performance analysis to the generalised multi-user, multi-PA scenario. Since the effective channel is formed by the superposition of multiple PA-level contributions with different in-waveguide attenuation, phase variations, and wireless propagation distances, its exact distribution becomes analytically intractable. To enable tractable analysis, the PA locations are modelled as independently distributed along the waveguide, and the corresponding PA-level channel contributions are treated as independent. Based on this approximation, the LoS communication and sensing channels are characterised using a moment-matching approach, whereas the NLoS communication channels are characterised using a PA-conditioned semi-analytical approach. Subsequently, the outage probability and ergodic sum-rate analyses of the proposed multi-PA PASS-QRSMA-ISAC framework are presented.
\subsection{\textbf{Statistical Characterization of Effective Channels}}
\label{ChannelCharacterisation}
\subsubsection{\textbf{Channel Characterization of CUs}}
The outage and ergodic sum-rate analyses in the multi-PA scenario require the statistical characterisation of the effective channels. From \eqref{CU_Channel}, the effective channel is obtained by summing the contributions from multiple PAs. Since the PA-level terms depend on the PA locations through the in-waveguide attenuation, phase variation, and wireless propagation distance, the exact distribution of the resulting effective channel is difficult to obtain in closed form. Therefore, we first condition on the location of the considered CU and characterise the randomness induced by the PA locations\footnote{For analytical tractability, the PA locations are modelled as independently distributed without explicitly imposing a hard minimum-spacing constraint. In practical scenarios, a minimum inter-PA spacing can be imposed to mitigate coupling, and a separation of $\frac{\lambda}{2}$ is commonly considered \cite{10981775}. For the considered system parameters, imposing this spacing constraint only removes a limited subset of closely spaced PA configurations and does not materially alter the overall PA-to-node distance distribution or the resulting channel statistics. Hence, its impact on the evaluated performance is negligible.}. For the CUs in Room 1, a moment-matching approach is used to approximate the effective channel as a complex Gaussian random variable. For the CUs in Room 2, the channel is characterised by conditioning on the CU and PA locations, after which the Rayleigh fading can be averaged. The corresponding spatially averaged performance is then obtained by integrating over the node locations.
\par The statistical characterization of the CU effective channel is performed by conditioning on a fixed CU location $\Phi_k$. However, obtaining the exact distribution of the resulting sum in closed form is analytically intractable due to the coupled dependence of amplitude and phase on the propagation distances. To enable tractable analysis, the effective summed channel of any user in Room 1 conditioned on $\Phi_k$ can be expressed as $h_{\Sigma,k}\sim\mathbb{CN}\{\mu_k(\Phi_k),\varsigma_k^2(\Phi_k)\}$ where the conditional mean and conditional variance for $k\leq M$ can be respectively expressed as:
\begin{equation}
    \mu_k(\Phi_k)=L\mathbb{E}\{h_{\omega,l}h_{k,l}|\Phi_k\}=\frac{L}{D}\int_{-D/2}^{D/2}h_{\omega,l}(r)h_{k,l}(r)dr,
\end{equation}
\begin{equation}
\begin{split}
    &\varsigma_k^2(\Phi_k)=\\&L\left[\frac{1}{D}\int_{-D/2}^{D/2}\left\lvert h_{\omega,l}(r)h_{k,l}(r)\right\rvert^2dr-\left\lvert \frac{1}{D}\int_{-D/2}^{D/2} h_{\omega,l}(r)h_{k,l}(r) dr\right\rvert^2\right],
\end{split}
\end{equation}
where $r$ denotes the generic PA position along the waveguide evaluated at $\Phi_{P_l}=[r,0,d]$. Now, let us consider the effective channel gain as $H_k=\lvert h_{\Sigma,k}\rvert^2$. For $k\leq M$, $H_k$ follows a non-central chi-square distribution with two degrees of freedom. Thus, the conditional PDF and CDF of $H_k$ can be respectively given as:
\begin{equation}
    f_{H_k}(x|\Phi_k)=\frac{1}{\varsigma_k(\Phi_k)^2}e^{-\frac{x+\lvert\mu_k(\Phi_k) \rvert^2}{\varsigma_k(\Phi_k)^2}}I_0\left(\frac{2\lvert\mu_k(\Phi_k) \rvert\sqrt{x}}{\varsigma_k(\Phi_k)^2}\right),
\end{equation}
\begin{equation}
    \EuScript{F}_{H_k}(x|\Phi_k)=1-\mathbb{Q}_1\left(\frac{\sqrt{2}\lvert\mu_k(\Phi_k)\rvert}{\varsigma_k(\Phi_k)},\frac{\sqrt{2x}}{\varsigma_k(\Phi_k)}\right),
\end{equation}
where $I_0(.)$ is the modified Bessel function of first kind of order zero, and $\mathbb{Q}_1(.,.)$ denotes the first order Marcum-Q function.
\par Now, for the CUs located in Room 2, i.e. $M+1\leq k\leq K$, the effective channel contains both the PA-location-dependent in-waveguide channel and the Rayleigh small-scale fading and can be expressed as $h_{\Sigma,k}=\sum\nolimits_{l=1}^{L}h_{\omega,l}g_{k,l}\left\lVert \Phi_{P_l}-\Phi_k \right\rVert^{-\nu/2}$. Now, conditioned on the CU location $\Phi_k$ and the PA locations $\{\Phi_{P_l}\}_{l=1}^L$, the terms $h_{\omega,l}\left\lVert \Phi_{P_l}-\Phi_k \right\rVert^{-\nu/2}$ become deterministic, while $g_{k,l}$ remains random. Therefore, the conditioned effective channel is a zero-mean complex Gaussian random variable, given by
\begin{equation}
    h_{\Sigma,k}\Big\lvert \Phi_k,\{\Phi_{P_l}\}_{l=1}^L\sim\mathbb{CN}\left\{0,\widehat{\varsigma}^2\left(\Phi_k,\{\Phi_{P_l}\}_{l=1}^L\right)\right\},
\end{equation}
where 
\begin{equation}
    \widehat{\varsigma}^2\left(\Phi_k,\{\Phi_{P_l}\}_{l=1}^L\right)=\sum_{l=1}^L\lvert h_{\omega,l}\rvert^2\left\lVert \Phi_{P_l}-\Phi_k \right\rVert^{-\nu},\quad M+1\leq k\leq K.
\end{equation}
Consequently, defining $H_k=\lvert h_{\Sigma,k}\rvert^2$, the conditional CDF of $H_k$ for the CUs in Room 2 can be expressed as:
\begin{equation}
    \EuScript{F}_{H_k}\left(x\lvert\Phi_k,\{\Phi_{P_l}\}_{l=1}^L\right)=1-e^{\frac{-x}{\widehat{\varsigma}^2\left(\Phi_k,\{\Phi_{P_l}\}_{l=1}^L\right)}}.
\end{equation}
This expression is used for the PA-conditioned semi-analytical evaluation of the NLoS CU outage probability, where the Rayleigh fading is averaged in closed form after conditioning on the CU and PA locations. The remaining expectation over the CU and PA locations is evaluated numerically in the subsequent outage analysis.
\begin{remark}
    For the CUs in Room 1, the effective channel is mainly governed by deterministic LoS propagation phases and PA locations. Hence, the moment-matched Rician approximation captures both the mean and variance of the summed channel. In contrast, for the CUs in Room 2, the effective channel contains both PA-location randomness and Rayleigh small-scale fading. If the PA-location-dependent channel power is first replaced by its spatial average, the outage probability is evaluated using an averaged channel power rather than the actual PA-conditioned channel power. This may introduce a mismatch for finite values of $L$. To avoid this, we use a PA-conditioned semi-analytical evaluation, where the Rayleigh fading is averaged in closed form after conditioning on the PA locations, and the remaining spatial expectation is evaluated numerically.
\end{remark}
\subsubsection{\textbf{Channel Characterization for Sensing}} Similar to Room 1 CU channels, the transmit side sensing channel as shown in \eqref{Tx_Sense_Channel} is characterised by conditioning on a fixed ST location $\Phi_S$. Accordingly, using moment matching, $h_{\Sigma,S}$ can be approximated as $h_{\Sigma,S}\sim\mathbb{CN}\{\mu_S(\Phi_S),\varsigma_S^2(\Phi_S)\}$ where the conditional mean and variance can be respectively given as:
\begin{equation}
    \mu_S(\Phi_S)=L\mathbb{E}\{h_{\omega,l}h_{S,l}|\Phi_S\}=\frac{L}{D}\int_{-D/2}^{D/2}h_{\omega,l}(r)h_{S,l}(r)dr,
\end{equation}
\begin{equation}
\begin{split}
    &\varsigma_S^2(\Phi_S)\\&=L\left[\frac{1}{D}\int_{-D/2}^{D/2}\left\lvert h_{\omega,l}(r)h_{S,l}(r)\right\rvert^2dr-\left\lvert \frac{1}{D}\int_{-D/2}^{D/2} h_{\omega,l}(r)h_{S,l}(r) dr\right\rvert^2\right].
\end{split}
\end{equation}
Defining $H_{\Sigma,S}=\lvert h_{\Sigma,S}\rvert^2$, the conditional CDF can be given by
\begin{equation}
\label{CDF_Tx_Sensing}
    \EuScript{F}_{H_{\Sigma,S}}(x|\Phi_S)=1-\mathbb{Q}_1\left(\frac{\sqrt{2}\lvert\mu_S(\Phi_S)\rvert}{\varsigma_S(\Phi_S)},\frac{\sqrt{2x}}{\varsigma_S(\Phi_S)}\right).
\end{equation}
The echo return channel as shown in \eqref{Rx_Echo_Channel} is characterised separately since the echo is collected only through the selected receiving PA which is closest to the ST. Thus, defining $\widehat{H_{\Sigma,S}}=\left\lvert \widehat{h_{\Sigma,S}}\right\rvert^2$, we have $\widehat{H_{\Sigma,S}}=\left\lvert \widehat{h_{\Sigma,S}}\right\rvert^2=10^{-\frac{\kappa\left\lVert \Phi_{\widehat{P_l}}-\Phi_{F} \right\rVert}{10}}\left\lVert \Phi_{S}-\Phi_{\widehat{P_l}} \right\rVert^{-2}$. From \eqref{overall_sensing_channel}, the effective sensing channel power gain is $H_{sen}=\lvert h_{sen}\rvert^2=\xi_SH_{\Sigma,S}\widehat{H_{\Sigma,S}}$. Now, for a fixed $\Phi_S$ and a selected receiving PA coordinate $\Phi_{\widehat{P_l}}$, the return-side channel gain $\widehat{H_{\Sigma,S}}$ is deterministic. Hence, using \eqref{CDF_Tx_Sensing}, the conditional CDF of $H_{sen}$ can be written as:
\begin{equation}
\label{HSEN}
    \begin{split}
    \EuScript{F}_{H_{sen}}\left(x\lvert\Phi_S,\Phi_{\widehat{P_l}}\right)&=\Pr\left(H_{sen}\leq x\lvert \Phi_S,\Phi_{\widehat{P_l}}\right)\\
    &=\Pr\left(\xi_SH_{\Sigma,S}\widehat{H_{\Sigma,S}}\leq x\lvert \Phi_S,\Phi_{\widehat{P_l}}\right)\\
    &=\Pr\left(H_{\Sigma,S}\leq \frac{x}{\xi_S\widehat{H_{\Sigma,S}}}\Bigg\lvert \Phi_S,\Phi_{\widehat{P_l}}\right)\\
    &=\EuScript{F}_{H_{\Sigma,S}}\left(\frac{x}{\xi_S\widehat{H_{\Sigma,S}}}\Bigg|\Phi_S\right).\\
    \end{split}
\end{equation}
Since $\widehat{P_l}$ is selected as the PA closest to the ST, the distribution of $x_{\widehat{P_l}}$ is induced by the nearest PA selection rule. Let us consider $\hat{r}=x_{\widehat{P_l}}$. Since the PAs are deployed only along the x-axis, the closest-PA selection depends only on the horizontal separation $\lvert \hat{r}-x_S\rvert$, while $y_S$ affects only the return-side channel gain. Thus, for a given $x_S$, the conditional PDF of the selected PA location can be expressed as:
\begin{equation}
    f_{x_{\widehat{P_l}}}(\hat{r}\lvert x_S)=\frac{L}{D}\left(1-\frac{\EuScript{W}(\hat{r}, x_S)}{D}\right)^{L-1},
\end{equation}
where $\EuScript{W}(\hat{r}, x_S)=\min\left(\frac{D}{2},x_S+\lvert \hat{r}-x_S\rvert\right)-\max\left(\frac{-D}{2},x_S-\lvert \hat{r}-x_S\rvert\right)$ represents the length of the interval around the ST within which another PA would be closer than the PA located at $\hat{r}$. Hence, the spatially averaged CDF of $H_{sen}$ is given by:
\begin{equation}
\label{HSen_Characterisation}
\begin{split}
    &\EuScript{F}_{H_{sen}}(x)=\\&\frac{1}{D^2}\int_{\frac{-D}{2}}^{\frac{D}{2}}\int_{\frac{-D}{2}}^{\frac{D}{2}}\int_{\frac{-D}{2}}^{\frac{D}{2}}\EuScript{F}_{H_{\Sigma,S}}\left(\frac{x}{\xi_S\widehat{H_{\Sigma,S}}(\hat{r},\Phi_S)}\Bigg|\Phi_S\right)f_{x_{\widehat{P_l}}}(\hat{r}\lvert x_S)d\widehat{r}dy_Sdx_S,
\end{split}
\end{equation}
where $\widehat{H_{\Sigma,S}}=10^{-\frac{\kappa\left\lvert \hat{r}-x_{F} \right\rvert}{10}}\left( (\hat{r}-x_S)^2+y_S^2+d^2 \right)^{-1}.$

\subsection{\textbf{Outage Analysis}}
In this sub-section, we provide the user COP, system COP and SOP analysis, respectively for the proposed multi-user multi-PA framework. 
\subsubsection{\textbf{COP Analysis}}
\label{COP_Section_multiplePA}
Due to the Q-RSMA-based transmission, each CU decodes the common stream from the in-phase component and its intended private stream from the quadrature component. Any CU is said to be in outage if either its common or private stream is in an outage. Thus, using $H_k=\lvert h_{\Sigma,k}\rvert^2$ and equations \eqref{Common_SINR} and \eqref{Private_SINR}, the COP of $U_k$ can be expressed as:
\begin{equation}
\begin{split}
    \text{P}_{Ok}&=1-\Pr\left\{\Lambda_{k,Com,Pin}>\tau_{Ck}, \Lambda_{k,Pvt,Pin}>\tau_{Pk}\right\}=\Pr\left\{H_k\leq\delta_k\right\},
\end{split}
\end{equation}
where $\delta_k=\max\left(\delta_{Ck},\delta_{Pk}\right)$ and
\begin{subequations}
    \label{Threshold_Uk}
    \begin{align}
        \delta_{Ck}(\tau_{Ck})&=\frac{L\tau_{Ck}\sigma^2}{\EuScript{P}_t\left(2\alpha\beta_\text{C}-\tau_{Ck}(1-\alpha)\right)},\label{Threshold_Common_Uk}\\
        \delta_{Pk}(\tau_{Pk})&=\frac{L\tau_{Pk}\sigma^2}{\EuScript{P}_t\left(2\alpha\beta_{Pk}-\tau_{Pk}(2\alpha\sum_{\substack{i=1, i\neq k}}^K\beta_{Pi}+1-\alpha)\right)}\label{Threshold_Pvt_Uk}.
    \end{align}    
\end{subequations}
Here, $\tau_{Ck}=2^{2\mathbb{R}_{Ck,Th}}-1$ and $\tau_{Pk}=2^{2\mathbb{R}_{Pk,Th}}-1$. Here, $\mathbb{R}_{Ck,Th}$ and $\mathbb{R}_{Pk,Th}$ are the target data rates of the common and private streams of $U_k$, respectively. Furthermore, if $\left(2\alpha\beta_\text{C}-\tau_{Ck}(1-\alpha)\right)\leq0$ or $\left(2\alpha\beta_{Pk}-\tau_{Pk}(2\alpha\sum_{\substack{i=1, i\neq k}}^K\beta_{Pi}+1-\alpha)\right)\leq0$, then $\text{P}_{Ok}=1$. Using the statistical characterization developed in Section \ref{ChannelCharacterisation}, the conditional COP at a given user location $\Phi_k$ can be given as:
\begin{equation}
\text{P}_{Ok}(\Phi_k)=
    \begin{cases}
        1-\mathbb{Q}_1\left(\frac{\sqrt{2}\lvert\mu_k(\Phi_k)\rvert}{\varsigma_k(\Phi_k)},\frac{\sqrt{2\delta_k}}{\varsigma_k(\Phi_k)}\right), & k\leq M\\
        \mathbb{E}_{\{\Phi_{P_l}\}_{l=1}^L}\left[1-e^{-\frac{\delta_k}{\widehat{\varsigma}_k^2\left(\Phi_k,\{\Phi_{P_l}\}_{l=1}^L\right)}}\right], & M< k\leq K.
    \end{cases}
\end{equation}
Now, since the CUs are uniformly distributed inside their respective rooms, the unconditional per-user COP is obtained by averaging the conditional outage probability over the corresponding user-location region which can be expressed as:
\begin{equation}
\begin{split}
    &\text{P}_{Ok}=\EuScript{F}_{H_k}(\delta_k)\\&=
        \begin{cases}
            \frac{1}{D^2}\int_{\frac{-D}{2}}^{\frac{D}{2}}\int_{\frac{-D}{2}}^{\frac{D}{2}}1-\mathbb{Q}_1\left(\frac{\sqrt{2}\lvert\mu_k(x_k,y_k)\rvert}{\varsigma_k(x_k,y_k)},\frac{\sqrt{2\delta_k}}{\varsigma_k(x_k,y_k)}\right)dx_kdy_k, & k\leq M\\
            \frac{1}{D^2}\int_{\frac{D}{2}}^{\frac{3D}{2}}\int_{\frac{-D}{2}}^{\frac{D}{2}}\mathbb{E}_{\{\Phi_{P_l}\}_{l=1}^L}\left[1-e^{-\frac{\delta_k}{\widehat{\varsigma}_k^2\left(\Phi_k,\{\Phi_{P_l}\}_{l=1}^L\right)}}\right]dy_kdx_k, & k>M
        \end{cases}
        \end{split}
\end{equation}
where $\mathbb{E}_{\{\Phi_{P_l}\}_{l=1}^L}$ denotes the spatial averaging over the PA locations, where $x_{P_l}\in\left[\frac{-D}{2},\frac{D}{2}\right]$.\\
$\bullet$ \textbf{System Outage:} Based on the discussion in Section \ref{singlePA_analysis}, the system outage as presented in \eqref{system_outage_singlaPA}, can be extended to the multi-PA analysis and can be given as:
\begin{equation}
\label{SystemOutage_MultiPA}
    \text{P}_{Sys}\approx1-\prod_{k=1}^K(1-\text{P}_{Ok}).
\end{equation}
Since the CUs located in the same room are assumed to follow the same spatial distribution, channel model, target-rate requirements, and power allocation rule, their outage probabilities are statistically identical after averaging over the user locations and channel fading. Let the spatially averaged outage probabilities of representative CUs in Room 1 and Room 2 be denoted by $\bar{\text{P}}_{O,R1}$ and $\bar{\text{P}}_{O,R2}$, respectively. Accordingly, for $M$ CUs in Room 1 and $K-M$ CUs in Room 2, the system outage probability can be approximated as:
\begin{equation}
\label{ApproxSysOutage}
    \text{P}_{Sys}\approx1-\left((1-\bar{\text{P}}_{O,R1})^M(1-\bar{\text{P}}_{O,R2})^{K-M}\right).
\end{equation}
\subsubsection{\textbf{SOP Analysis}}
\label{SOP_Section_multiplePA}
Similar to the single-PA case, the CRLB is considered for target-range estimation, and hence the same sensing-SINR threshold $\tau_S$ is used. Thus, following the MSE-based SOP definition in Section \ref{SOP_Analysis_SinglePA}, the multi-PA SOP can be given as:
\begin{equation}
\label{MultiPA_SensingOutage}
\begin{split}
    \text{P}_{OS}&=\Pr(\Lambda_{Sensing,Pin}\leq \tau_S)\\&=\Pr\left(H_{sen}\leq \frac{L\tau_S\sigma_S^2}{\EuScript{P}_t\left[(1-\alpha)-\alpha\tau_S\Delta_S\right]}\right)\\&=\EuScript{F}_{H_{sen}}\left(\frac{L\tau_S\sigma_S^2}{\EuScript{P}_t\left[(1-\alpha)-\alpha\tau_S\Delta_S\right]}\right),
\end{split}
\end{equation}
where $\EuScript{F}_{H_{sen}}(.)$ can be found in \eqref{HSen_Characterisation}. Owing to the nearest-PA selection and the spatial averaging over the ST location, the final SOP expression is evaluated using the trapezoidal rule. It should be noted that if $\left[(1-\alpha)-\alpha\tau_S\Delta_S\right]\leq0$, then $\text{P}_{OS}=1$.
\subsection{\textbf{Ergodic Sum Rate Analysis}}
\label{ESRA}
In this sub-section, the ergodic sum-rate of the considered multi-PA framework is evaluated using the channel-gain distributions derived in Section \ref{ChannelCharacterisation}.
\begin{equation}
\label{SumRate_Ana_Multple Users}
\begin{split}
    \EuScript{R}_{\Sigma,Ana}&=\mathbb{E}\left\{\min(\EuScript{R}_{\text{C}1},\ldots,\EuScript{R}_{\text{C}K})\right\}+\sum_{k=1}^K\mathbb{E}\left\{\EuScript{R}_{\text{P}k}\right\}.
\end{split}
\end{equation}
Using the CCDF-based identity discussed in Section \ref{singlePA_analysis}, the average common-rate component can be approximated as:
\begin{equation}
\label{COMMON_RATE_ANALYSIS}
    \mathbb{E}\left\{\EuScript{R}_C\right\}\approx\frac{1}{2\ln 2}\int_0^\infty \frac{\prod_{k=1}^K \left[1-\EuScript{F}_{H_k}(\delta_{Ck}(x))\right]}{1+x}dx,
\end{equation}
and the average private rate component can be approximated as:
\begin{equation}
\label{PVT_RATE_ANALYSIS}
    \mathbb{E}\left\{\EuScript{R}_{Pk}\right\}\approx\frac{1}{2\ln 2}\int_0^\infty \frac{\left[1-\EuScript{F}_{H_k}(\delta_{Pk}(x))\right]}{1+x}dx,
\end{equation}
where $\delta_{Ck}(x)$ and $\delta_{Pk}(x)$ are obtained from \eqref{Threshold_Common_Uk} and \eqref{Threshold_Pvt_Uk}, by replacing $\tau_{Ck}$ and $\tau_{Pk}$ with $x$. The integrals in \eqref{COMMON_RATE_ANALYSIS} and \eqref{PVT_RATE_ANALYSIS} are evaluated numerically using the trapezoidal rule.
\section{Results and Discussion}
\label{results}
\begin{table}[t]
\centering
\caption{Simulation Parameters \cite{ShivSir,cheng2025performance}.}
\label{tab:PARAMETERS}
\renewcommand{\arraystretch}{1.25}
\begin{tabular}{|c||c|}
 \hline
 \rowcolor{green!20} {\bf Parameters} & {\bf Values}\\
 \hline
 \rowcolor{blue!20}\multicolumn{2}{|c|}{\textbf{General System Parameters}} \\
  \hline
  Transmit Power, $\EuScript{P}_t$ & $-20 \text{ to } 30$ dBm\\
  \hline
  Carrier Frequency, $f_C$ & 10 GHz\\
  \hline
  Length of each room, $D$ & $10-20$ m\\
  \hline
  Height at which waveguide is placed, $d$ & $5-20$ m\\
  \hline
  ISAC power splitting factor, $\alpha$ & 0.7\\
  \hline
  Background Noise Power, $\sigma^2=\sigma_S^2$ & $10^{-12}$ W\\
  \hline
  Number of PAs, $L$ & 1-10\\
  \hline
  Effective refractive index of waveguide, $\eta_{eff}$ & 1.4\\
  \hline
  In-waveguide attenuation coefficient, $\kappa$ & 0-0.2 dB/m\\
  \hline
  Number of Monte Carlo Simulations & $10^6$\\
  \hline
  \rowcolor{blue!20}\multicolumn{2}{|c|}{\textbf{Communication Parameters}} \\
  \hline
  Total number of CUs, $K$ & 2-10\\
  \hline
  Number of CUs in Room 1, $M$ & $\frac{K}{2}$\\
  \hline
  Number of CUs in Room 2 & $K-M$\\
  \hline
  Path loss exponent, $\nu$ & 6\\
  \hline
  Quadrature points, $\EuScript{M}$ & 75\\
  \hline
  Q-RSMA Power allocation factor, $\left\{\beta_\text{C}, \beta_\text{P}k\right\}$& $\left\{0.6, \frac{1-\beta_\text{C}}{K}\right\}$\\
  \hline
  Target data rates, $\{\mathbb{R}_{Ck,Th}, \mathbb{R}_{Pk,Th}\}$ & $\{0.7, 0.2\}$\\
  \hline
  \rowcolor{blue!20}\multicolumn{2}{|c|}{\textbf{Sensing Parameters}} \\
  \hline
  Radar Cross-Sectional Area, $\sigma_{RCS}$ & $\sim0-20$ m$^2$\\
  \hline
  Residual interference coefficient, $\Delta_S$ & 0-0.05\\
  \hline
  CRLB parameter, $k_S$ & $\frac{64\pi^2}{12}$\\
  \hline 
  Target MSE Threshold, $\rho_{Th}$ & 0.2, 0.5, 0.8\\
  \hline
  Bandwidth Fraction ratio, $\varpi$ & [0,1]\\
  \hline
  Bandwidth, $\EuScript{B}$ & 15 MHz\\
  \hline
\end{tabular}
\end{table}
In this section, the performance of the proposed PASS-QRSMA-ISAC framework is evaluated to demonstrate its operational behaviour and effectiveness. The simulation results are presented and compared with the corresponding analytical results to validate the accuracy of the developed analysis. It can be observed that the simulation results closely match the analytical results in all the presented figures, thereby verifying the accuracy of the derived analytical expressions for the per-user COP, system COP, sum rate, and SOP. In addition, the proposed framework is compared with traditional frameworks like PASS-RSMA-ISAC, PASS-NOMA-ISAC, and PASS-SDMA-ISAC to highlight its performance advantages. Specifically, the PASS-RSMA-ISAC benchmark follows the model in \cite{ShivSir}, whereas the PASS-NOMA-ISAC benchmark follows the model in \cite{cheng2025performance}. It should be noted that these existing benchmark models were originally developed for the single-PA, two-user scenario. Hence, they are first adopted in their original form for the single-PA, two-user comparison. For the generalised multi-PA and multi-user case, the benchmark schemes are extended following the same PASS-assisted channel modelling and signal construction principles discussed in \cite{10945421}, so that all the schemes are evaluated under a consistent system setting. Furthermore, the PASS-SDMA-ISAC benchmark is obtained from the PASS-RSMA-ISAC model by setting the power allocated to the common stream to zero, such that only the private streams are transmitted. In the figures, the simulation results are denoted by \say{Sim}, while the analytical results are represented by \say{Ana}. The system parameters used throughout the evaluation are summarized in Table \ref{tab:PARAMETERS}.
\subsection{\textbf{Communication Task Performance}}
In this subsection, the communication performance of the proposed PASS-QRSMA-ISAC framework is evaluated. The proposed framework is compared with PASS-RSMA-ISAC, PASS-NOMA-ISAC, and PASS-SDMA-ISAC. Since only the communication components of these frameworks are evaluated in this subsection, they are hereafter referred to as QRSMA-PASS, RSMA-PASS, NOMA-PASS, and SDMA-PASS, respectively.
\begin{figure}[t]
    \centering

    \begin{subfigure}{0.49\linewidth}
        \centering
        \includegraphics[scale=0.42]{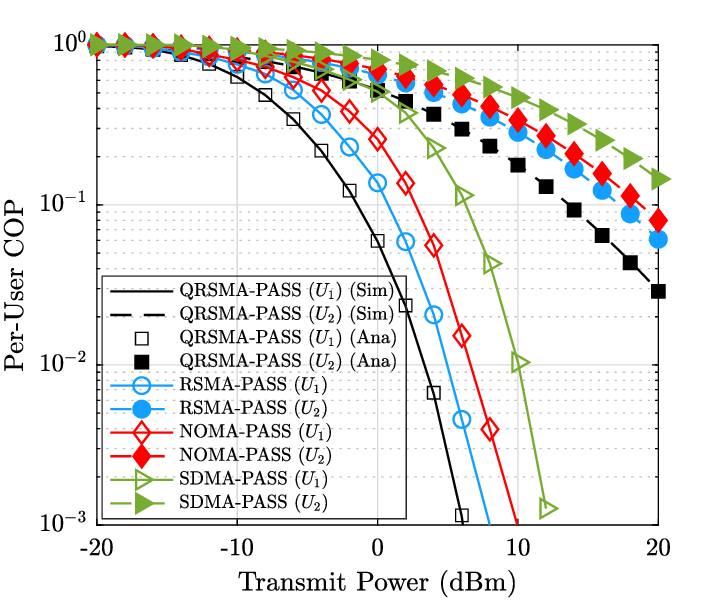}
        \caption{$\EuScript{P}_t$ v/s per-user COP performance for two-user single-PA scenario.}
        \label{SinglePA_COP_Power}
    \end{subfigure}
    \hfill
    \begin{subfigure}{0.49\linewidth}
        \centering
        \includegraphics[scale=0.42]{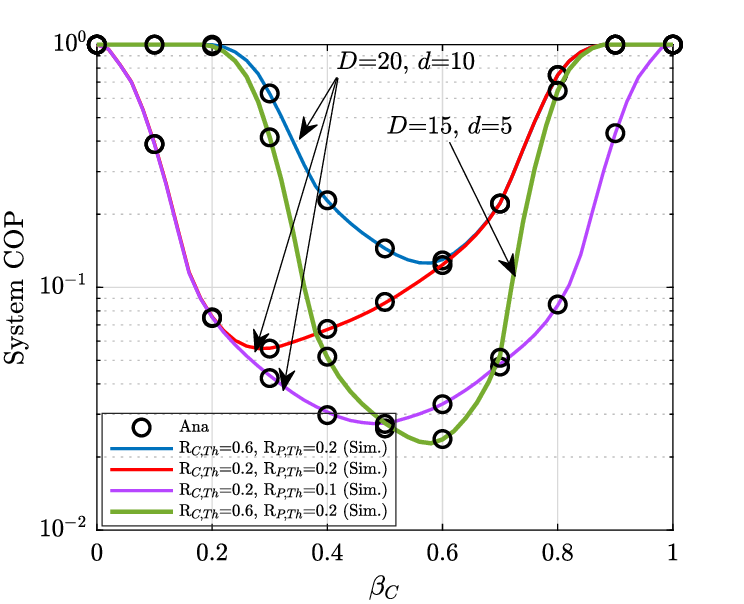}
    \caption{$\EuScript{P}_t$ v/s system COP performance for two-user single-PA scenario.}
    \label{SinglePA_SystemCOP_Power}
    \end{subfigure}
    \caption{(a) Per-user COP performance comparison of the proposed framework with other traditional frameworks wrt $\EuScript{P}_t$, and (b) system COP performance evaluation for varying target thresholds wrt common stream power allocation factor, $\beta_C$. $\left(\text{R}_{C,Th}=\mathbb{R}_{Ck,Th} \text{ and } \text{R}_{C,Th}=\mathbb{R}_{Ck,Th}\right)$}
    \label{SinglePA_COP}
\end{figure}
\par \figurename{~\ref{SinglePA_COP}}(a) illustrates the per-user COP of the proposed framework and benchmark schemes with respect to the transmit power, $\EuScript{P}_t$. It can be observed that the COP of both users decreases as $\EuScript{P}_t$ increases, since higher transmit power improves the received SINR and enhances the decoding reliability. However, $U_2$ which is located in Room 2, exhibits a higher per-user COP than $U_1$. This is because $U_2$ experiences NLoS propagation and fading, whereas $U_1$ benefits from a stronger LoS link with the PA. Among the considered schemes, the proposed QRSMA-PASS framework achieves the best per-user COP performance for both users. This improvement arises from the interference-free transmission of the common stream along with the inherent rate-splitting gains, which enhance decoding reliability. PASS-RSMA provides the second-best performance and outperforms both PASS-NOMA and PASS-SDMA due to the benefits of rate-splitting, which enable more efficient interference management. In contrast, PASS-SDMA exhibits the worst performance, even inferior to PASS-NOMA. This behaviour occurs because the considered system is overloaded, i.e., the number of transmit pinching antennas is smaller than the number of users. In such scenarios, spatial multiplexing alone cannot effectively separate the users, which leads to significant residual inter-user interference. Since PASS-SDMA treats this interference purely as noise without employing interference management or successive interference cancellation mechanisms, it results in a higher outage probability compared with the other considered schemes.
\par \figurename{~\ref{SinglePA_COP}}(b) illustrates the system COP wrt the common-stream power allocation factor, $\beta_C$, for different target thresholds at $\EuScript{P}_t$=15 dBm. It can be observed that the system COP first decreases with $\beta_C$, reaches a minimum, and then increases again. When $\beta_C$ is small, insufficient power is allocated to the common stream, and hence the common-stream decoding reliability at both users becomes poor. As $\beta_C$ increases, the common-stream SINR improves, which reduces the system COP. However, after a certain point, further increasing $\beta_C$ reduces the power available for the private streams. Consequently, the private-stream decoding reliability deteriorates, and the system COP starts increasing again. Therefore, an optimal $\beta_C$ exists that balances the decoding requirements of the common and private streams. The effect of the target thresholds can also be observed from the figure. When the common-stream threshold is reduced while keeping the private-stream threshold fixed, reliable common-stream decoding can be achieved with a smaller value of $\beta_C$. As a result, the optimal operating point shifts leftward. In contrast, when the private-stream threshold is reduced while keeping the common-stream threshold fixed, the private streams require less power for successful decoding. Hence, a larger fraction of power can be allocated to the common stream, which shifts the optimal point rightward. Furthermore, reducing the room dimension and waveguide deployment height from $D$=20 m and $d$=10 m to $D$=15 m and $d$=5 m improves the overall outage performance due to shorter propagation distances and stronger LoS links. However, for a given set of target thresholds, the optimal $\beta_C$ remains almost unchanged, since the relative trade-off between the common and the private stream power allocations is mainly governed by the decoding thresholds rather than the absolute channel strength.
\begin{figure}[t]
    \centering

    \begin{subfigure}[t]{0.49\linewidth}
        \centering
        \includegraphics[scale=0.42]{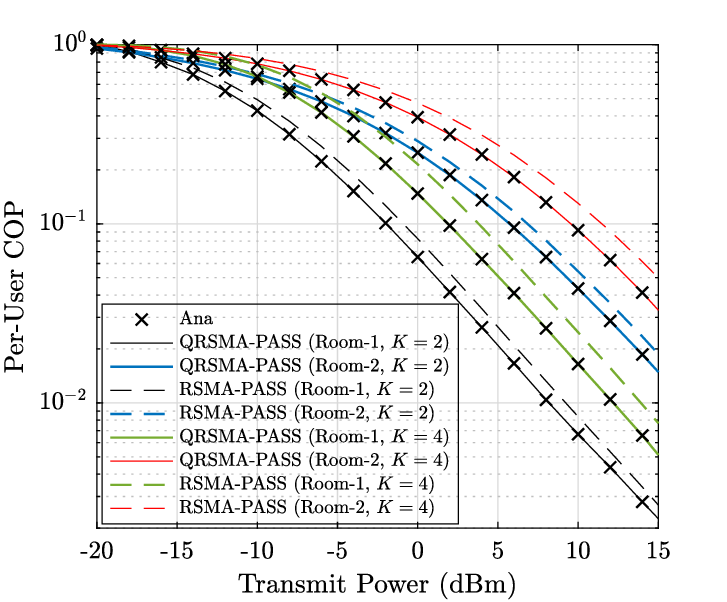}
        \caption{Per-user COP performance comparison wrt $\EuScript{P}_t$ for different values of $K$.}
        \label{MultiPA_COP_Power}
    \end{subfigure}
    \hfill
    \begin{subfigure}[t]{0.49\linewidth}
        \centering
        \includegraphics[scale=0.42]{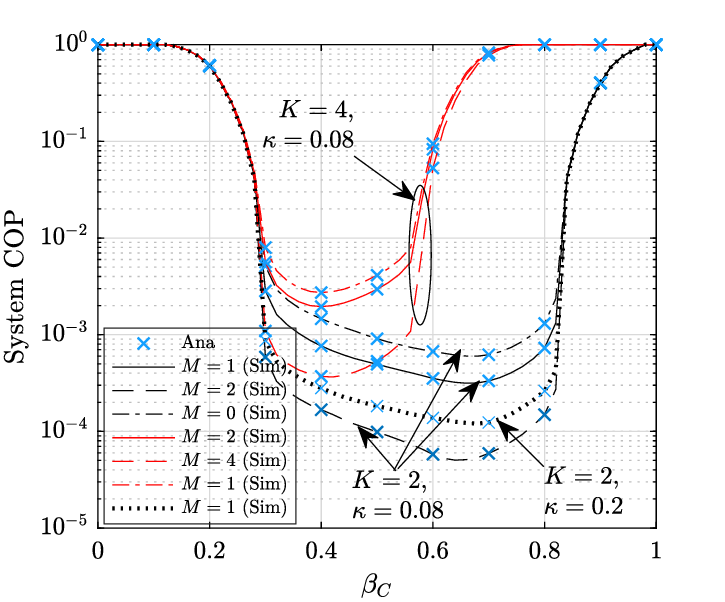}
    \caption{System COP performance evaluation wrt $\beta_C$. $(\mathbb{R}_{Ck,Th}=0.5, \mathbb{R}_{Pk,Th}=0.12)$}
    \label{MultiPA_SystemCOP_Power}
    \end{subfigure}

    \caption{COP performance comparison of the multi-user multi-PA scenario for $L$=5.}
    \label{MultiPA_COP}
\end{figure}
\par \figurename{~\ref{MultiPA_COP}}(a) illustrates the per-user COP performance of the proposed QRSMA-PASS framework and the RSMA-PASS framework wrt $\EuScript{P}_t$, for different values of $K$ in the multi-user multi-PA scenario. Since the users located in the same room follow the same spatial distribution, channel model, target-rate requirements, and power-allocation rule, their outage behaviour is statistically identical after averaging over user locations and channel fading. Therefore, one representative user from each room is selected to illustrate the per-user COP behaviour. It can be observed that the per-user COP decreases with increasing $\EuScript{P}_t$ for all considered cases, since higher transmit power improves the received SINR and enhances decoding reliability. Moreover, users located in Room 1 achieve better COP performance than those located in Room 2. This is because Room 1 users experience stronger LoS links with the PAs, whereas Room 2 users are affected by NLoS propagation and fading, resulting in weaker channel conditions. It can also be observed that QRSMA-PASS outperforms RSMA-PASS across all values of $K$. This improvement is due to the signal-domain separation in QRSMA, where the common stream and private streams are transmitted over orthogonal signal components, thereby reducing the interference affecting common-stream decoding. In contrast, RSMA-PASS relies on SIC-based stream separation, which makes its decoding reliability more sensitive to residual interference. Furthermore, increasing $K$ degrades the COP performance for both frameworks, since a larger number of users increases the number of interfering private streams and reduces the power available for each private stream. Consequently, the decoding conditions become more restrictive, leading to higher COP.
\par \figurename{~\ref{MultiPA_COP}}(b) illustrates the system COP wrt $\beta_C$, for different values of $K$, $M$, and $\kappa$ in the multi-user multi-PA scenario, where $\EuScript{P}_t$=20 dBm, $D$=10 m, and $d$=5 m. Since the users located in the same room follow the same spatial distribution, channel model, target-rate requirements, and power-allocation rule, their outage behaviour is statistically identical after averaging over user locations and channel fading. Hence, the spatially averaged COP of one representative user from each room is used to obtain the system COP according to \eqref{ApproxSysOutage}. It can be observed that the system COP initially decreases with $\beta_C$, reaches a favourable operating point, and then increases again. This behaviour is due to the trade-off between the common and the private stream decoding requirements. When $\beta_C$ is small, insufficient power is allocated to the common stream, resulting in poor common-stream decoding reliability. As $\beta_C$ increases, the common-stream SINR improves, thereby reducing the system COP. However, beyond a certain point, the power available for the private streams decreases, which degrades private-stream decoding and causes the system COP to increase. It can also be observed that increasing the value of $K$ degrades the system COP and shifts the favourable operating region towards smaller values of $\beta_C$. This is because a larger number of users reduces the power allocated to each private stream and also increases the number of interfering private streams. Consequently, the private-stream decoding condition becomes more restrictive, and the system cannot allocate a large fraction of power to the common stream without significantly degrading the private-stream reliability. Hence, the favourable $\beta_C$ region shifts leftward for larger $K$. Furthermore, the value of $M$ has a noticeable impact on the system COP. For a fixed value of $K$, increasing $M$ implies that more users are located in Room 1 and fewer users are located in Room 2. Since Room 1 users experience LoS links with the PAs, whereas Room 2 users experience NLoS propagation, a larger $M$ generally improves the system COP performance. In contrast, smaller $M$ leads to a larger number of NLoS users, which weakens the overall system reliability and increases the system COP. Finally, increasing $\kappa$ worsens the system COP performance because a larger in-waveguide attenuation coefficient affects the received SINR of both the common and private streams decreases, leading to a higher probability of outage.
\begin{figure}[t]
    \centering

    \begin{subfigure}[t]{0.49\linewidth}
        \centering
        \includegraphics[scale=0.42]{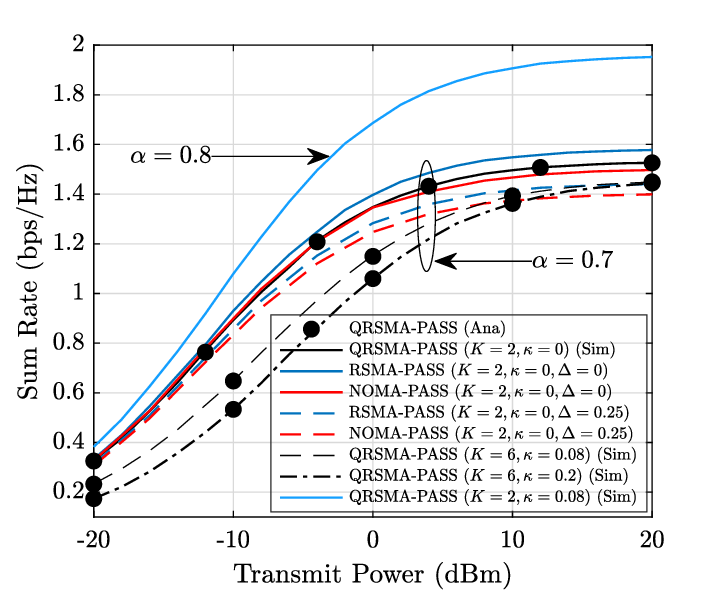}
        \caption{$\EuScript{P}_t$ v/s sum rate.}
        \label{MultiPA_Power_SumRate}
    \end{subfigure}
    \hfill
    \begin{subfigure}[t]{0.49\linewidth}
        \centering
        \includegraphics[scale=0.42]{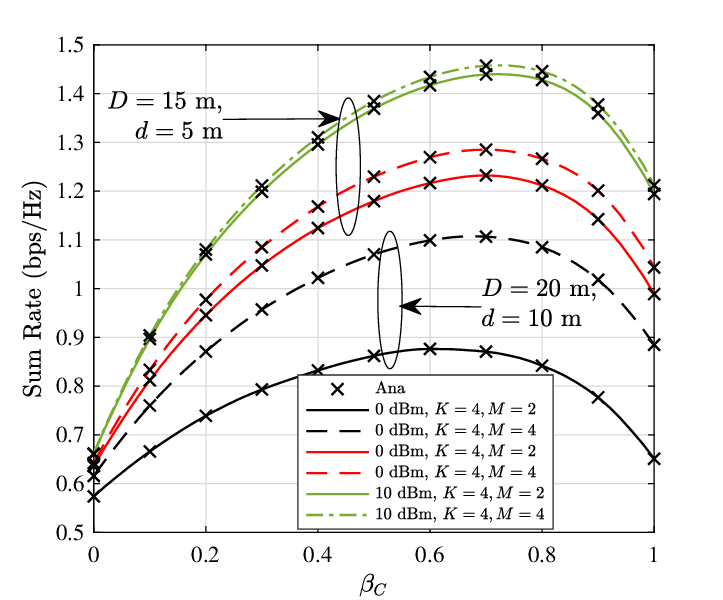}
    \caption{$\beta_C$ v/s sum rate.}
    \label{MultiPA_PowerCoeff}
    \end{subfigure}
    \caption{(a) Sum rate performance of the proposed QRSMA-PASS framework with benchmark RSMA-PASS and NOMA-PASS frameworks wrt $\EuScript{P}_t$ for different values of $K$, $\kappa$ and $\Delta$. ($\Delta:$ ipSIC parameter), and (b) $\beta_C$ v/s sum rate for different values of $\EuScript{P}_t$, $M$, $D$ and $d$.}
    \label{MultiPA_SumRate}
\end{figure}

\par \figurename{~\ref{MultiPA_SumRate}}(a) illustrates the sum-rate performance of the proposed QRSMA-PASS framework wrt $\EuScript{P}_t$, in comparison with the RSMA-PASS and NOMA-PASS benchmark schemes. To evaluate practical decoding conditions, the impact of imperfect SIC is considered for RSMA-PASS and NOMA-PASS, where the ipSIC parameter is denoted using $\Delta$. Here, we have considered ipSIC due to hardware constraints. The simulation results for the sum rate have been generated using \eqref{SumRate_Ana_Multple Users} while the analytical results have been obtained using the equations as discussed in Section \ref{ESRA}. It can be observed that the sum rate increases with $\EuScript{P}_t$ for all the considered cases. This is because a higher transmit power improves the received SINR of the decoded streams, thereby increasing the achievable rate. However, at higher transmit power, the curves tend to saturate. This occurs because both the desired communication signal and the sensing-signal interference increase with $\EuScript{P}_t$, which makes the SINR interference-limited rather than noise-limited. Moreover, when $\alpha$ is increased from 0.7 to 0.8, the overall sum rate increases, and the near-linear growth region extends over a wider transmit power range before the curves approach saturation. This is because, a larger $\alpha$ allocates more power to the communication signal while reducing the power assigned to the sensing signal. Consequently, the desired communication signal strength increases, whereas the sensing-signal interference decreases, thereby delaying the transition towards the interference-limited region. For QRSMA-PASS, the use of real-valued signalling over the in-phase and quadrature components introduces a scaling factor of $\frac{1}{2}$ in the achievable rate expressions. Therefore, this signal-domain separation reduces the available DoF and can affect the sum-rate performance, especially when the interference level is not dominant. This is observed in the figure as QRSMA-PASS exhibits a lower sum-rate performance compared to RSMA-PASS under perfect SIC conditions. Now, when compared to NOMA-PASS, QRSMA-PASS exhibits a higher sum-rate performance at higher transmit power owing to the rate-splitting gain and improved interference management although the effects of reduced DoF is still visible at lower transmit power where NOMA-PASS exhibits better sum rate than QRSMA-PASS. Under practical conditions with $\Delta>0$, the advantage of QRSMA-PASS becomes evident, particularly at higher transmit power. This is because RSMA-PASS and NOMA-PASS rely on SIC-based decoding and therefore suffer from residual interference under imperfect SIC. In contrast, QRSMA-PASS separates the common and private streams over orthogonal signal components and does not require SIC between these streams. Hence, it is more robust to SIC imperfections and achieves better sum-rate performance in the practical high-power region. Furthermore, increasing $K$ reduces the sum-rate performance under fixed total transmit power. This is because a larger number of users reduces the power available for each private stream and increases the number of interfering private streams, thereby degrading the private-stream SINRs. In addition, increasing $\kappa$ reduces the sum rate, since the signal experiences stronger in-waveguide attenuation before being radiated through the PAs, which weakens the effective channel gain and reduces the received SINR.
\par \figurename{~\ref{MultiPA_SumRate}}(b) illustrates the sum-rate performance of the proposed QRSMA-PASS framework wrt $\beta_C$, for different values of $\EuScript{P}_t$, $M$, $D$ and $d$. It can be observed that the sum rate initially increases with $\beta_C$, reaches a maximum, and then decreases. This behaviour is due to the power-allocation trade-off between the common and private streams. When $\beta_C$ is small, the common stream receives limited power and its rate contribution is restricted. As $\beta_C$ increases, the common-stream SINR improves, and the gain in the common-rate contribution dominates the loss caused by the reduced private-stream power. However, beyond the optimal region, the reduction in private-stream power becomes dominant, which decreases the private-stream SINRs and their corresponding rate contributions. Consequently, the overall sum rate starts to decrease. Now, for a fixed value of $K$=4, increasing $M$ from 2 to 4 improves the sum rate, since more users are located in Room 1 and experience LoS-dominant links with the PAs. In contrast, when $M$=2, two users are located in Room 2 and experience NLoS propagation, which weakens the effective channel gains and reduces the achievable rate. Furthermore, the performance gap between the $M$=2 and $M$=4 cases is larger for $D$=20 m and $d$=10 m than for $D$=15 m and $d$=5 m. This is because larger room dimensions and PA height increase the propagation distances, making the Room 2 NLoS users more severely affected by path loss and fading. Hence, replacing Room 2 users with Room 1 users provides a more significant rate gain under the larger $D$ and $d$ setting. Finally, this gap reduces at higher transmit power. At lower $\EuScript{P}_t$, the weaker NLoS users are more power-limited, so the benefit of having more LoS users is more visible. As $\EuScript{P}_t$ increases, the received signal strength improves for both Room 1 and Room 2 users. However, the rate improvement becomes progressively limited because the interference terms also increase with transmit power, causing the system to move towards a saturation region. As a result, the additional gain obtained by replacing NLoS users with LoS users becomes less pronounced at higher transmit power. Consequently, the sum-rate gap between the $M=2$ and $M=4$ cases becomes smaller at higher transmit power.
\subsection{\textbf{Sensing Task Performance}}
In this sub-section, we present the SOP results for the two-user single-PA and multi-user multi-PA scenarios to evaluate the sensing performance of the proposed framework under different system parameters. The obtained results illustrate the sensing behaviour of the considered system and demonstrate the effectiveness of the proposed scheme in supporting reliable sensing functionality. For the two-user single-PA scenario, the simulated SOP results are obtained from \eqref{Sensing_SNR} by setting $L=1$, whereas the corresponding analytical results are derived using \eqref{SOP_Normal} and \eqref{SOP_CDF}. For the multi-user multi-PA scenario, the analytical SOP results are obtained using \eqref{HSEN} and \eqref{MultiPA_SensingOutage}. It can be observed that the simulated and analytical results closely match under both perfect and imperfect interference cancellation conditions, thereby validating the accuracy of the derived analytical expressions.
\begin{figure}[t]
    \centering
    \begin{subfigure}{0.49\linewidth}
        \centering
        \includegraphics[scale=0.42]{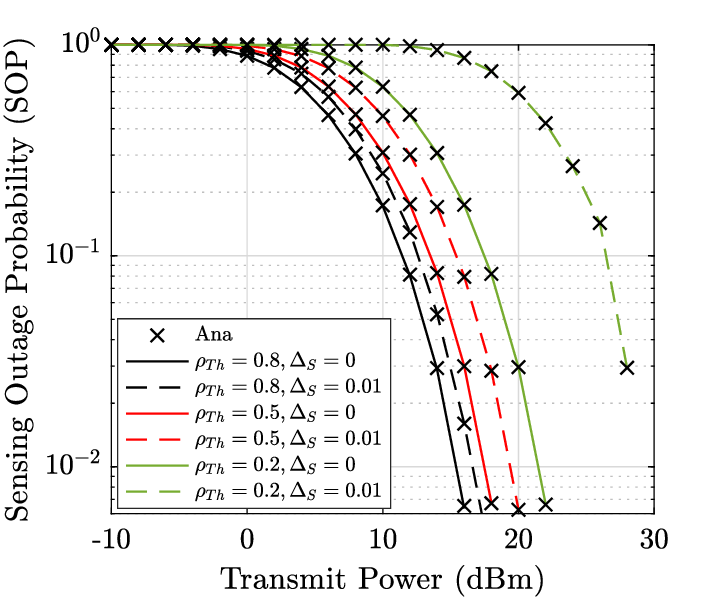}
        \caption{$\EuScript{P}_t$ v/s SOP with varying $\rho_{Th}$ and $\Delta_S$ for $L$=1, $\kappa$=0, and $D$=10 m.}
        \label{SinglePA_SOP_Power}
    \end{subfigure}
    \hfill
    \begin{subfigure}{0.49\linewidth}
        \centering
        \includegraphics[scale=0.42]{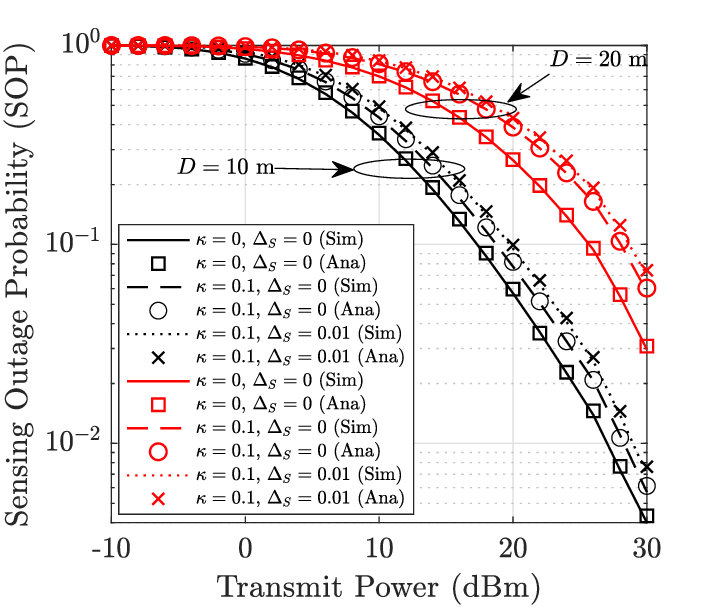}
    \caption{$\EuScript{P}_t$ v/s SOP for varying $\kappa$ and $\Delta_S$ for $L$=5 and $\rho_{Th}$=0.8.}
    \label{MultiPA_SOP_Power}
    \end{subfigure}
    \caption{Performance of sensing outage probability wrt $\EuScript{P}_t$.}
    \label{Sensing_Outage}
\end{figure}
\par \figurename{~\ref{Sensing_Outage}} illustrates the SOP performance of the proposed PASS-QRSMA-ISAC framework wrt $\EuScript{P}_t$. It can be observed from both the single-PA and multi-PA results that the SOP decreases as the transmit power increases. This is because a higher transmit power improves the received sensing SINR, thereby reducing the probability that the sensing SINR falls below the required threshold. In particular, \figurename{~\ref{Sensing_Outage}}(a) presents the SOP of the sensing target, $U_S$, for the two-user single-PA scenario under different target MSE thresholds i.e. $\rho_{Th}$=0.2, 0.5, and 0.8, considering both perfect and imperfect interference cancellation. The results show that a larger $\rho_{Th}$ leads to better SOP performance. This follows from the sensing threshold $\tau_S=\frac{c^2}{k_S\rho_{Th}(\varpi\EuScript{B})^2}$ which shows that the required sensing SINR is inversely proportional to $\rho_{Th}$. Hence, increasing $\rho_{Th}$ relaxes the sensing accuracy requirement, while a smaller $\rho_{Th}$ imposes a stricter sensing constraint. This confirms the trade-off between sensing accuracy and sensing reliability. Moreover, under imperfect interference cancellation, the SOP becomes higher than that under perfect cancellation because residual communication interference increases the effective interference level during sensing-signal recovery. This degradation is more pronounced for smaller $\rho_{Th}$, since the stricter sensing requirement makes the system more sensitive to residual interference. On the other hand, \figurename{~\ref{Sensing_Outage}}(b) presents the SOP performance for the generalised multi-PA scenario under ideal and practical waveguide conditions, corresponding to $\kappa=0$ and $\kappa=0.1$, respectively. It can be observed that an increase in the value of $\kappa$ worsens the SOP, since the signal experiences additional attenuation while propagating through the waveguide before being radiated by the PAs. The degradation further worsens when imperfect interference cancellation is considered, as both waveguide attenuation and residual interference jointly reduce the effective sensing SINR. Moreover, increasing the room dimension $D$ also worsens the SOP because the average propagation distance between the PAs and the sensing target increases, thereby weakening the round-trip sensing channel and reducing the received echo strength. This also explains why the impact of $\kappa$ becomes more pronounced for larger $D$, since a larger deployment region generally increases the effective in-waveguide propagation distance, causing the attenuation accumulated inside the waveguide to have a stronger effect on the received sensing echo.

\begin{figure*}[t]
\centering
\begin{minipage}{1\columnwidth}
    \centering
    \includegraphics[scale=0.5]{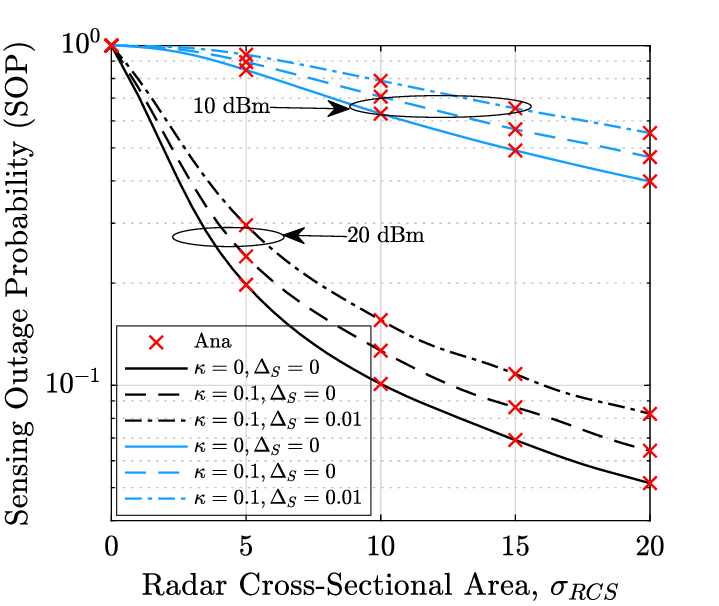}
    \caption{$\sigma_{RCS}$ v/s SOP for $L$=10, $D$=10 m, and $\rho_{Th}$=0.8 under different system configurations.}
    \label{SOP_RCS}
\end{minipage}
\begin{minipage}{1\columnwidth}
    \centering
    \includegraphics[scale=0.48]{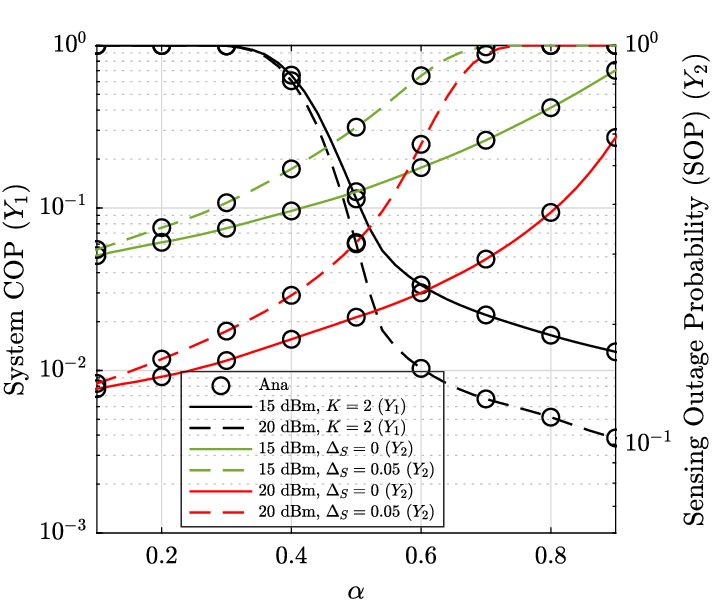}
    \caption{System COP ($Y_1$) and SOP ($Y_2$) wrt ISAC power splitting factor, $\alpha$.}
    \label{ISAC_Figure}
\end{minipage}
\vspace{-1.25em}
\end{figure*}
\par \figurename{~\ref{SOP_RCS}} illustrates the SOP performance with respect to the radar cross-sectional area, $\sigma_{RCS}$ for different values of $\EuScript{P}_t$, $\kappa$, and $\Delta_S$. As expected, the SOP decreases with increasing $\sigma_{RCS}$, since a larger radar cross-section strengthens the reflected echo and improves the sensing SINR. Now, it can be observed that the SOP is worse for $\kappa$=0.1 when compared to the ideal waveguide case in which $\kappa$=0. This is because a non-zero $\kappa$ introduces additional in-waveguide attenuation, which reduces the effective sensing channel gain and weakens the received echo signal. Moreover, the SOP further deteriorates when $\Delta_S>0$, since imperfect interference cancellation leaves residual communication interference during sensing-signal recovery. Therefore, the combined effect of waveguide attenuation and residual interference results in the worst SOP performance. Furthermore, the effects of both $\kappa$ and $\Delta_S$ become more visible at higher transmit power. At lower transmit power, the SOP curves remain closer to the high-outage region, limiting the visible separation among different practical impairment cases. However, at higher transmit power, the ideal case benefits more significantly from the stronger received echo, whereas the practical cases remain constrained by accumulated in-waveguide attenuation and residual interference. Hence, the SOP gaps caused by both $\kappa$ and $\Delta_S$ become more pronounced.
\vspace{-0.5em}
\subsection{\textbf{ISAC Performance}}
In this sub-section, we present the ISAC performance by jointly illustrating the system COP (shown on the $Y_1$ axis) and the SOP (shown on the $Y_2$ axis) wrt the power allocation factor, $\alpha$, as depicted in \figurename{~\ref{ISAC_Figure}}. The parameter $\alpha$ determines the fraction of the total transmit power allocated to the communication task, while the remaining portion (1-$\alpha$) is allocated to the sensing task. Now, when the value of $\alpha$ is small, most of the transmit power is assigned to sensing, leaving limited power for communication. As a result, the SINR of the communication streams (i.e the common and the private streams) become insufficient to satisfy the decoding thresholds, leading to a higher system COP. As $\alpha$ increases, the power allocated to communication gradually increases. Initially, the system COP does not change significantly since the communication SINR is still insufficient to meet the required thresholds. However, beyond a certain value of $\alpha$, the SINR corresponding to the common and the private streams becomes adequate to support successful decoding, causing the system COP to decrease. In contrast, the SOP exhibits an opposite trend. As $\alpha$ increases, the power allocated to the sensing task decreases since it depends on $(1-\alpha)\EuScript{P}_t$. Consequently, the received sensing SINR reduces, resulting in an increase in the SOP. Furthermore, when imperfect interference cancellation is considered, the SOP performance degrades compared to the perfect cancellation case. This degradation becomes more pronounced at higher values of $\alpha$ because the residual interference component increases with the communication power, which is proportional to $\alpha\EuScript{P}_t$. Therefore, the impact of imperfect interference cancellation becomes more significant at larger $\alpha$ values, which results in a wider gap between the perfect and imperfect cancellations plots.

\vspace{-0.5em}
\section{Conclusion}
\label{conclusion}
This paper proposed a PASS-QRSMA-ISAC framework for downlink communication and sensing, where multiple PAs deployed on a waveguide simultaneously serve multiple CUs and perform target sensing. The dual-room setup captured heterogeneous LoS/NLoS propagation conditions, and the corresponding communication and sensing signal models were developed along with the SNR/SINR expressions. A two-user single-PA case under ideal waveguide conditions was first analysed, where the per-user COP, system COP, SOP, and ergodic sum-rate were formulated using distance statistics, CDF/CCDF characterisations, and numerical integration methods. The analysis was then extended to the generalised multi-user multi-PA scenario by developing tractable statistical characterisations for the communication and sensing channels. Based on these characterisations, the per-user COP, system COP, SOP, and ergodic sum-rate were formulated and evaluated using numerical integration methods. Monte Carlo simulations validated the analysis and demonstrated the gains of Q-RSMA over the considered benchmark schemes within the PASS framework. The results further revealed that the communication-sensing trade-off is strongly governed by user distribution, power allocation, waveguide attenuation, and sensing-related parameters, since these factors determine the effective channel strength, interference level, and sensing recovery performance.
\par As part of our future work, we plan to extend the proposed framework by incorporating imperfect CSI and mobility-induced channel variations. Furthermore, an optimization-based design framework will be developed to improve the communication–sensing trade-off, including PA-position-based sum-rate maximization, adaptive power allocation, and dynamic PA activation. More advanced sensing tasks, such as joint range–angle estimation, target tracking, and multi-target sensing, may also be considered to enhance the practicality of PASS-QRSMA-ISAC systems.

\vspace{-1.5em}
\balance
\bibliographystyle{IEEEtran}
\bibliography{Bibliography.bib}
\end{document}